\documentclass[aps,prb,twocolumn,groupedaddress,showpacs,floatfix,altaffilletter]{revtex4-1}
\usepackage{graphicx}
\usepackage{grffile}
\usepackage{amsmath}
\usepackage{amssymb}
\usepackage{bm}
\usepackage{color}
\usepackage[dvipsnames]{xcolor}
\usepackage{amsmath,amssymb,amsfonts}
\usepackage{epsfig}
\usepackage{times}
\usepackage[colorlinks,bookmarks=false,citecolor=blue,linkcolor=red,urlcolor=blue]{hyperref}

\newcommand{\fref}[1]{Fig.~\ref{#1}}
\newcommand{\eref}[1]{Eq.~(\ref{#1})}

\begin{document}

\title{Hall and Faraday effects in interacting multi-band systems with arbitrary band topology and spin-orbit coupling}

\author{R.~Nourafkan$^{1}$ and A.-M.S. Tremblay$^{1,2}$}
\affiliation{$^1$D{\'e}partement de physique, Institut quantique, Regroupement qu\'eb\'ecois sur les mat\'eriaux de pointe, Universit{\'e} de Sherbrooke, Sherbrooke, Qu{\'e}bec, Canada  J1K 2R1}
\affiliation{$^2$Canadian Institute for Advanced Research, Toronto, Ontario, Canada M5G 1Z8}
\date{\today}
\begin{abstract}
A formula for the Hall response of interacting multi-band systems with arbitrary band topology and spin-orbit coupling  is derived. The formula is valid at finite frequency, which is relevant for Faraday rotation, and it takes into account all particle-hole vertex corrections. The formula includes both three-leg (triangular) and four-leg (rectangular)  diagrams. The latter diagrams are absent in the single band case. We show that the rectangular diagrams are necessary to recover the semiclassical formula for the Hall effect from the Kubo formula in the DC limit. They also give the linear response of the anomalous Hall effect to an external magnetic field, an effect which goes beyond the semiclassical theory. Three- and four-particle scatterings are neglected. 

\end{abstract}

\maketitle
\section{Introduction}

The zero-frequency (DC) Hall conductivity is widely used as a probe of Fermi surface evolution~\cite{Nair_Wirth_Friedemann_Steglich_Si_Schofield_2012} and topological properties of eigenstates. Despite its long history ~\cite{Hall:1879}, the theory of the Hall effect in the presence of interactions and of multi-band effects, both at finite frequencies and in the DC limit, remains unattended. At finite frequency, the Hall effect controls Faraday rotation~\cite{PhysRevLett.70.2004} and is expected to contain important information, just as the optical conductivity does~\cite{RevModPhys.83.471}. Recent results in high-Tc superconductors,~\cite{Nature.10.1038/nature16983, PhysRevB.95.224517} heavy fermions~\cite{Nair_Wirth_Friedemann_Steglich_Si_Schofield_2012}, iron-based superconductors~\cite{PhysRevLett.109.096402}, topological insulators~\cite{Hall_topology_Luttinger}, Dirac and Weyl semimetals~\cite{Zhang_Narayan:2017} for example, show the continued importance of the Hall effect. 
%
%
%
%

In many studies of the Hall effect, the calculations are based on formulae that are direct extensions of the single-band formula. These extensions can be inadequate for multi-band systems because, as we will show, they consider only the three-leg (triangular) diagrams.~\cite{PhysRevB.45.13945, Prog.Theor.Phys.80.623} 

In this paper, we derive a formula for the Hall effect that includes both three-leg and four-leg diagrams. The formula is valid for model Hamiltonians that describe multi-band systems in the presence of arbitrary interactions. It includes vertex corrections: While the interaction of an excited electron or hole with the environment is taken into account through the self-energy, the interaction of an excited electron with the accompanying excited hole is taken into account through vertex corrections to the current operator. This has to be taken into account to understand responses of interacting systems, whether the calculation is for model Hamiltonians of for realistic materials calculations that use density-functional methods merged with many-body techniques~\cite{Anisimov_Poteryaev_Korotin_Anokhin_Kotliar_1997,KotliarRMP:2006,Pavarini:17645,Aichhorn_Pourovskii:2015,Galler_Kaufmann_Gunacker_Thunstrom_Tomczak_Held_2017}. 
For example, Boltzmann transport theory predicts a small Hall effect for a multi-band system with several hole and electron pockets once electrons and holes have comparable densities and mobilities. This situation occurs for iron-based superconductors but measured Hall effects are very large with a material dependent carrier character.  This discrepancy was explained, in the framework of a simple extension of the Boltzmann transport theory to multi-band case, by taking into account vertex correction effects due to dominant interband interactions~\cite{PhysRevLett.109.096402} which mix bands. As we show here, the band mixing effects are present even at the bare level if the correct formula for the multi-band case is used.

We first discuss in Sec.~\ref{Sec:Motivation} the current vertices that are needed for model Hamiltonians, the case that we consider here. The coupling to the electromagnetic field due to currents circulating between atoms requires a discussion of the Peierls substitution. There are ambiguities with that substitution that are highlighted and resolved. Appendix~\ref{app:AFM} illustrates the ambiguities in the simple case of the commensurate antiferromagnet. In the frequency-dependent case, the Peierls substitution does not suffice to describe interband transitions that are driven by intra-atomic currents. Electric-dipole matrix elements must be taken into account. A sketch of the derivation for the Hall effect in the general case and our final formula are in Sec.~\ref{Sec:Hall_Response}. The details of the derivation are in Appendices~\ref{Appendix_A} and ~\ref{D}. Appendix~\ref{Appendix_A} in particular, addresses some subtle issues related to vertex corrections. We show in Sec.~\ref{Sec:DC_Semiclassical} how one recovers the semiclassical limit. Appendix~\ref{app:AFM} demonstrates for the antiferromagnetic state of the one-band Hubbard model, why the rectangular diagrams can be important to obtain the correct semi-classical formula when there is a valence and a conduction band. The conclusions are in Sec.~\ref{Sec:Conclusions} while Appendix~\ref{Sec:Appendix_DC} discusses some details related to the link to the semiclassical theory and Appendix~\ref{Sec:Matsubara} contains various Matsubara frequency summations that we need for derivations. We restrict ourselves to a discussion of the longitudinal and Hall conductivities, but similar arguments are valid for the analog thermoelectric quantities, thermopower, Nernst coefficient and thermal Hall (Righi-Leduc) effect. 
\section{Coupling to the electromagnetic fields}\label{Sec:Motivation}
Minimal coupling in first-principles calculation and in model Hamiltonians require separate discussions. In the first subsection below, we define what is a model Hamiltonian in the orbital basis and discuss how the electromagnetic field must be introduced in such a Hamiltonian to obtain the correct current and effective mass vertices. The second subsec- tion gives a simple example with physical discussion. The rest of this paper will be based on the approach described here.

\subsection{Minimal coupling in model Hamiltonians: the orbital basis}
Minimal coupling between matter and electromagnetic field leads to the following general prescription for the current operator as a functional derivative of the Hamiltonian $\hat{H}$ with respect to the vector potential $\mathbf{A}$:
\begin{align} \label{eq:dHdA}
\mathbf{\hat{J}(r)}=-\frac{\delta \hat{H}}{\delta \mathbf{A(r)}}\,.
\end{align}
In the simple case where there is no spin-orbit coupling, one recovers the well known second-quantized expression for the current operator
\begin{align} \label{eq:current_general}
\mathbf{\hat{J}(r)}=\frac{-e}{2m}\sum_{\sigma } \left[\psi^\dagger_\sigma(\mathbf{r})(\frac{\hbar}{i}\nabla+e\mathbf{A})\psi_\sigma(\mathbf{r})\right.\nonumber\\
\left. -((\frac{\hbar}{i}\nabla-e\mathbf{A})\psi^\dagger_\sigma(\mathbf{r})) \psi_\sigma(\mathbf{r})\right]
\end{align}
in terms of the elementary charge $e$ and of the field annihilation (creation) operators $\psi^{(\dagger)}_\sigma(\mathbf{r})$ for an electron of spin $\sigma$ at position $\mathbf{r}$.  In a weakly correlated crystalline system, the field operators are expanded in terms of  the Bloch functions.  

In an interacting system, it is more suitable to expand the field operators in terms of an atomic-like orbital basis set. In this case the  model Hamiltonians are constructed, in principle, by keeping only a few bands near the Fermi level and projecting them into  states that are orthonormal and localized around atomic sites (Wannier basis set). These states are labeled by $\mathbf{R}_i+\mathbf{r}_a$, that we call the position of the atomic site $a$ in unit cell $\mathbf{R}_i$, and by another label, $l$. Fourier transforming from the unit cell positions $\mathbf{R}_i$ to crystal momentum $\mathbf{k}$ one obtains what we call {\it the Hamiltonian in the orbital basis}. It is worth mentioning that there is no unique way to define the orthonormal localized basis since one can do a $\mathbf{k}$-dependent unitary transformation amongst the Bloch states. The results should in principle be independent of the choice of unitary transformation but, in practice, since we work in a truncated basis, some choices are better than others when modeling interactions~\cite{RevModPhys.84.1419}. Note that even though the labels $l$ do not, in general, correspond to specific orbitals, we will continue to refer to $l$ as an orbital index for simplicity.   

In the localized basis, there are two contributions to the current operator: (i) radiative transitions between atomic states and (ii) tunneling between different atomic sites. For the latter contribution it is preferable to start with the Hamiltonian in Wannier basis and preform the so-called Peierls substitution~\cite{Paul_Kotliar:2003}. Specifically, each of the terms in the kinetic-energy operator that represent hopping between atomic sites is multiplied by a phase 
\begin{align} 
    \exp\left[\frac{ie}{\hbar}\int_{\mathbf{R}_i+\mathbf{r}_a}^{\mathbf{R}_j+\mathbf{r}_b} \mathbf{A}\cdot d\mathbf{r}\right]\, ,
\end{align}
where the integral is along a straight line connecting the positions of the atoms between which hopping occurs, {\it even within a unit cell}. This is the only way to couple the electromagnetic field to a few given localized orbitals in a gauge-independent way. 

Because we loose translational invariance when the vector potential corresponding to a magnetic field is applied, the Hamiltonian is not diagonal in crystal momenta. Since we will be working in the long wavelength limit, one can find the current operator \eref{eq:dHdA} corresponding to a given Fourier component of the vector potential from 

\begin{align} \label{eq:dHorbdA}
    \mathbf{\hat{J}(q)}=-\sum_{\mathbf{k}\mathbf{k'}}\sum_{\sigma\sigma'}\sum_{l,l'}
    c_{\mathbf{k}\sigma l}^\dagger
    \frac{\partial H^{\mathbf{(A)} {\sigma l,\sigma' l'} }_{\mathbf{k,k'}}}{\partial \mathbf{A(q)}}
    c_{\mathbf{k'}\sigma'l'}\,.
\end{align}
with $c_{\mathbf{k}\sigma l}^{(\dagger)}$ standing for annihilation (creation) operators in the single-particle state with quantum numbers $\mathbf{k}\sigma l$. In the calculations below, $\mathbf{H^{(A)}}$ stands for the matrix that appears above, but without explicit indices.  With spin-orbit coupling, this matrix is not necessarily diagonal in spin indices. Its derivative with respect to vector potential leads to the current operator for both interband and intraband transitions.   

\subsection{A simple example and warnings}
In momentum space and for {\it the single-band case}, the general result \eref{eq:dHorbdA} for the current operator corresponding to a uniform electric field leads to an expression to the uniform ($\mathbf{q}=0$) gauge-invariant current due to hopping that, to leading order, is of the form
\begin{align} \label{eq:j_with_derivative}
\hat{J}^{\alpha}_{hop}=-\frac{e}{\hbar}\sum_{\mathbf{k}\sigma}\left( \frac{\partial E_{\mathbf{k}}}{\partial k_\alpha}+\frac{e}{\hbar}\frac{\partial^2 E_{\mathbf{k}}}{\partial k_\alpha^2}A_\alpha\right) c_{\mathbf{k}\sigma}^\dagger c_{\mathbf{k}\sigma} 
\end{align} 
where $\alpha$ labels spatial direction, $E_{\mathbf{k}}$ is the dispersion relation and $c_{\mathbf{k}\sigma}^{(\dagger)}$ is the annihilation (creation) operator for a state of wave vector $\mathbf{k}$ with spin $\sigma$. For simplicity of the present discussion, we assume that there is no spin-orbit interaction and drop the spin index on the dispersion relation.  

The above result for the current is based on a replacement of the derivative with respect to vector potential by a derivative with respect to wave vector $\mathbf{k}$, as suggested by the Peierls substitution.  It is important however to note that this replacement by a derivative with respect to $\mathbf{k}$ is a basis-dependent statement. In a multi-orbital case with more than one atom per unit cell, the correct minimal coupling is ${\it not}$ equivalent to the change of variables $E_{n,\mathbf{k}}\rightarrow E_{n,\mathbf{k}+e\mathbf{A}/\hbar}$, where $n$ is the band index. This is discussed further in Appendix~\ref{Sec:AFM}.

The physical choice of current vertex is the one obtained from a derivative with respect to $\mathbf{k}$ of the model Hamiltonian in the orbital basis (See \eref{current2} for an example). Indeed, take the case of a solid with two atoms, $A$ and $B$,  per unit cell. The bands are linear combinations of operators on the $A$ and $B$ sublattices that are separated in space. Hence, currents between these two sites are associated with a dipole that can lead to interband transition and they are taken into account by the Peierls substitution in the orbital basis. In fact, using the complete expression for the current, \eref{eq:current_general}, there are interband transitions in the general case, even in the band basis, as shown in Ref.~\onlinecite{doi:10.7566/JPSJ.84.124708}. The interband transitions come from wave-vector derivatives of the periodic part of the Bloch functions. 

Note that we can replace the derivative with respect to vector potential by a derivative with respect to ${\bf k}$ only when we neglect intra-atomic dipole transitions, do the Peierls substitution in the orbital basis and include the phase factor 
for hopping within the unit cell as well as between unit cells.  Otherwise, if the phase factor includes only the position of the unit cell, the current cannot be expressed as a derivative with respect to ${\bf k}$ only.~\cite{PhysRevB.80.085117}


\section{Longitudinal and Hall conductivities}\label{Sec:Hall_Response}
The Hall response is linear in applied electric field, hence linear response theory in Matsubara frequency can be used. The magnetic field by itself does not lead to a dissipative response. It can be included in the Hamiltonian to arbitrary order. We thus first recall the linear response formalism and then motivate the result for the Hall conductivity. Analytic continuation is performed as the last step of the calculation. Details of the derivation are given in Appendices~\ref{Appendix_A} and~\ref{D}. 

We choose a gauge where the electric field is represented by a time-dependent and space-independent vector potential, ${\bf A}(t)$, while the magnetic field is represented by a time-independent and space dependent vector potential, ${\bf A}'({\bf r})$. In other words,
\begin{align}
{\bf A}(t)+{\bf A}'({\bf r})&={\bf A }\exp[{-i\nu t}]+{\bf A}^\prime\exp[{i\mathbf{q}\cdot \mathbf{r}]}.
\end{align}
In Fourier space, we thus have for the electric and magnetic fields respectively ${\bf E}({\bf q},\nu)=i\nu{\bf A}({\bf q},\nu)$ and 
${\bf B}({\bf q}',\nu') =i{\bf q}'\times {\bf A}'({\bf q}',\nu')$.

Note that matrices in the orbital basis and vectors, such as the vector potential and the crystal momentum, are all represented indistinctly by bold symbols, even though they are not objects in the same space. The context will insure that no confusion arises.

\subsection{Conductivity in linear response}

We first consider the conductivity in zero magnetic field. The {\it expectation value} of the current operator \eref{eq:dHorbdA} in the $\alpha$ direction in the presence of the  time-dependent vector potential ${\bf A}$ that represents the electric field, is given in imaginary time by
\begin{equation}\label{eq:J_LambdaG_before_expanding}
J^{\alpha}(\tau)=\rm{Tr}\left[\left(-\partial_{A_\alpha}\mathbf{H}^{({\bf A})}\right)\mathbf{G}^{({\bf A})}(\tau,\tau+0^+)\right]
\end{equation} 
where we used the shortcut $\partial_{A_\alpha}$ for the partial derivative with respect to vector potential of the Hamiltonian matrix in the orbital basis \eref{eq:dHorbdA}, and where the expectation of the field operators with the non-equilibrium density matrix is now represented by a time-ordered Matsubara Green's function matrix in the same basis
\begin{equation}
    \mathbf{G}^{({\bf A})}(\tau,\tau')=-< T_\tau \mathbf{c}(\tau) \mathbf{c}^\dagger(\tau') >_{{\bf A}}\, .
\end{equation}
The trace in \eref{eq:J_LambdaG_before_expanding} is over crystal momentum, spin and band indices. Since the interaction does not depend on the vector potential, it is only the non-interacting part of $\mathbf{H}^{({\bf A})}$ that will be relevant.

We are interested in linear response. This means that we must expand both the vertex $-\partial_{A_\alpha}\mathbf{H}^{({\bf A})}$ and the Green's function $\mathbf{G}^{({\bf A})}$ to linear order in the vector potential, which gives what are known, respectively, as the diamagnetic and the paramagnetic contributions. The final answer for the uniform $\mathbf{q}=0$ result takes the general form
\begin{equation}
	J^{\alpha}(\nu)=\sum_{\beta}\Pi_{\alpha \beta}(\nu) A^{\beta}(\nu),  
\end{equation}
where  $\nu$ denotes the frequency.
Expanding in a basis that takes advantage of translational invariance by one unit cell, the Green's function ${\bf G}_{{\bf k}}$ becomes a $\mathbf{k}$ dependent matrix. 
In the Matsubara representation, the calculation is relatively simple. Making use of the matrix identity  
\begin{equation}\label{eq:dG=-GdG-1G}
\partial_{A_{\alpha}}{\bf G}^{({\bf A})}=-{\bf G}^{({\bf A})}\partial_{A_{\alpha}}{\bf G}^{({\bf A})-1}{\bf G}^{({\bf A})},
\end{equation}   
that follows from differentiating ${\bf G}^{({\bf A})}{\bf G}^{({\bf A})^{-1}}={\bm 1}$, one recovers the standard expression~\cite{Mahan} for the polarization tensor $\Pi_{\alpha \beta}(i\nu_n)$, namely
\begin{widetext}
\begin{align}
\Pi_{\alpha \beta}(i\nu_n)=-& \frac{\delta_{\alpha \beta}k_BT}{V_{\rm cell}N}\sum_{{\bf k}\omega_m}{\rm Tr} \left\{\left[
{\bm \lambda}_{\alpha\beta}({\bm 0},0)\right]_{{\bf k},\omega_m;{\bf k},\omega_m}{\bf G}_{{\bf k},\omega_m}\right\}  \nonumber\\ -&
\frac{k_BT}{V_{\rm cell}N}\sum_{{\bf k}\omega_m}{\rm Tr}\left\{\left[ {\bm \lambda}_{\alpha}({\bf 0},i\nu_n)\right]_{{\bf k},\omega_m;{\bf k},\omega_m^+}
{\bf G}_{{\bf k},\omega_m^+}
\left[{\bm \Lambda}_{\beta}({\bf 0},-i\nu_n)\right]_{{\bf k},\omega_m^+;{\bf k},\omega_m}
{\bf G}_{{\bf k},\omega_m}\right\},\label{Pi1}
\end{align}
\end{widetext}
where $T$ is the temperature, $k_B$ is Boltzmann's constant, $V_{cell}$ is the volume of the unit cell, $N$ is the number of unit cells, and all bold quantities are square matrices labeled with orbital and spin indices. These matrices are diagonal in spin indices only if there is no spin-orbit interaction. The traces from now on will be in orbital and spin indices (The sum over crystal momentum will be written explicitly). In the above equation, $\delta_{\alpha \beta}$ is the Kronecker delta, $i\nu_n$ is a bosonic Matsubara frequency, the bare current vertex is 
\begin{align}\label{eq:bare_vertex}
{\bm \lambda}_{\alpha}\equiv -\partial_{A_\alpha}{\bf H}^{({\bf A})}|_{{\bf A}={\bm 0}}
\end{align}
with ${\bf H}^{({\bf A})}$ the non-interacting part of Hamiltonian in the orbital basis, while the dressed current vertex in the same basis is
\begin{align}\label{eq:dressed_vertex}
{\bm \Lambda}_{\alpha}\equiv \partial_{A_{\alpha}}{\bf G}^{({\bf A})-1}|_{{\bf A}={\bm 0}}.
\end{align}
These vertices include both intra- and inter-atomic transitions and in the case of model Hamiltonians where the Peierls substitution is used properly, they can be calculated from derivatives with respect to $\mathbf{k}$ when intra-atomic dipole transitions can be neglected, as discussed in the previous section. 
The derivative  of the current vertex with respect to the gauge potential is 
\begin{align}
{\bm \lambda}_{\alpha\beta}\equiv \partial_{A_\beta}{\bm \lambda}^{({\bf A})}_{\alpha}|_{{\bf A}={\bm 0}}.
\label{eq:bare_inv_effective_mass}
\end{align}
It plays the role of the bare inverse effective-mass tensor (at least in the one-band case) and it comes from differentiating the vector potential in the gauge invariant expression for the current. 
The corresponding dressed vertex ${\bm \Lambda}_{\alpha\beta}$ is defined in Appendix~\ref{Appendix_A}. 

The fully interacting single-particle Green's function entering \eref{Pi1} is
\begin{equation}
{\bf G}_{{\bf k},\omega_m}=[(i\omega_m+\mu){\bm 1}-{\bf H} ({\bf k})-{\bm \Sigma}({\bf k},i\omega_m)]^{-1}, 
\end{equation}
where  ${\bm \Sigma}$ is the electron self-energy, $\mu$ is the chemical potential and $\omega_m$ denotes the fermionic Matsubara frequencies.  
We have also defined $\omega_m^{+}\equiv \omega_m + \nu_n$. Note that ${\bm \Lambda}_{\sigma}^{\alpha}$ contains a contribution that is equal to the bare current vertex and a contribution that comes from a functional derivative of the self-energy, which is the vertex correction. Vertex corrections depend on two fermionic frequencies and one bosonic frquency, with corresponding wave vectors.

The first term in \eref{Pi1} is the diamagnetic response while second term in this equation is the paramagnetic one. Both terms are necessary to insure gauge invariance. Indeed, the conductivity tensor is defined as $J^{\alpha}(\nu)=\sum_{\beta}\sigma_{\alpha \beta}(\nu) E^{\beta}(\nu)$. Thus, once analytic continuation is performed, the $\mathbf{q}=0$ conductivity is given by the retarded response
\begin{equation}\label{eq:cond_real}
\sigma_{\alpha \beta}(\nu)=\frac{1}{i(\nu+i 0^+)}\Pi^R_{\alpha \beta}(\nu),
\end{equation}
with $0^+$ an infinitesimal positive number. Since gauge invariance implies that there can be no response to a frequency and momentum-independent vector potential, we must have $\Pi_{\alpha \beta}(\nu=0)=0$. This means that the diamagnetic part is equal to minus of the paramagnetic part at $\nu=0$. When the derivative with respect to vector potential can be replaced by a derivative with respect ot $\mathbf{k}$, this can easily shown by integration by part of the diamagnetic term. Alternatively, using current conservation and the definition of the correlation functions in terms of current commutators in  real-frequency, one confirms that this result is true in general.


\subsection{Hall response}

\begin{widetext}

\begin{figure*}
	\includegraphics[width=0.9\columnwidth]{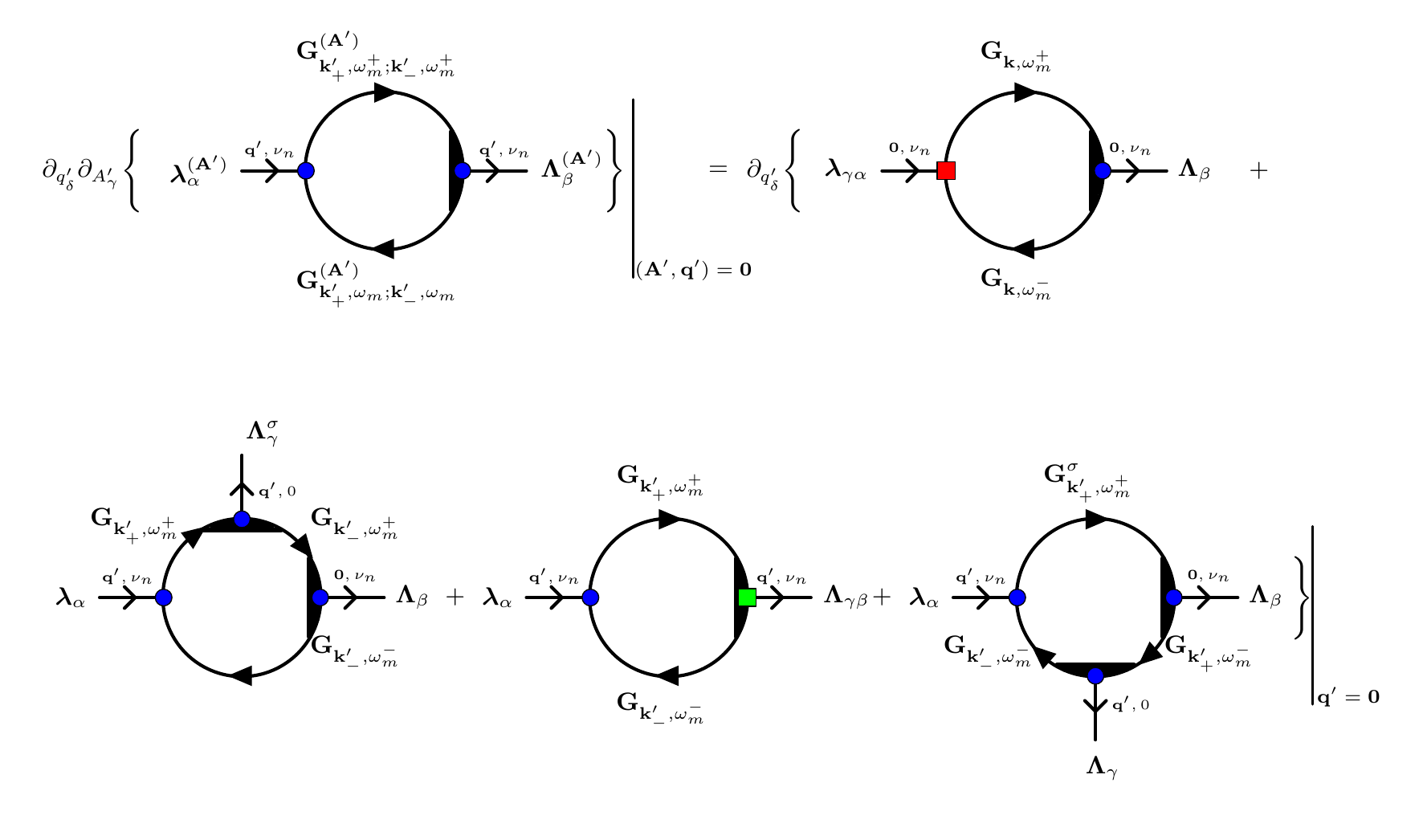}
	\caption{Diagrammatic expansion of the change in the current due to the presence of a magnetic field, evaluated in the zero-field limit. Lines show the fully dressed Green’s function, cirles and squares with their adjacent shaded area denote the dressed current vertex and its derivative, respectively. We also define ${\bf k}'_{\pm}\equiv {\bf k} \pm {\bf q}'/2$ and $\omega_m^{\pm} = \omega_m\pm\nu_n$.  Only diagrams at the second line contribute at the Hall response. The vertex $\mathbf{\Lambda}_{\gamma\beta}$, represented by an attached green square box, must be treated carefully, as discussed in Appendix~\ref{Appendix_A}. The final results include triangular (three-leg) and rectangular (four-leg) diagrams shown in the next two figures.}\label{fig:Diag}
\end{figure*}

\end{widetext}

In a time-reversal invariant, topologically trivial material, the transverse conductivity ($\alpha\neq \beta$) vanishes. When these conditions are not satisfied, because of a finite Berry curvature or a spontaneous magnetization for example, there may be a DC dissipationless intrinsic Hall effect that we obtain from the definition of the transverse conductivity \eref{eq:cond_real} as follows 
\begin{align}
\sigma_{\alpha \beta}^{intrinsic}=-i\lim_{\nu\rightarrow 0}\partial_{\nu}\Pi^R_{\alpha \beta}(\nu)\,\,\,\,;\,\,\,\, \alpha\ne\beta.
\end{align}

In general, the Hall response is the transverse conductivity ($\alpha\neq \beta$) in the presence of a perpendicular magnetic field, ${\bf B}$.  It thus suffices to compute the linear response of $\Pi_{\alpha \beta}$ to a magnetic field. The only analytic continuation we will need to do is that associated with $i\nu_n$ for the electric field since the magnetic field is time independent. We are thus justified to use the Matsubara formalism. We can represent a magnetic field $B^\eta=i\epsilon^{\eta\gamma\delta} q'_\gamma A'_\delta$ applied in a direction $\eta$ perpendicular to $\alpha,\beta$ by a vector potential written in a (Landau) gauge where $\mathbf{q'}$ is in a direction $\gamma$ in the $\alpha,\beta$ plane and $\mathbf{A'}$ is in the same plane but in a direction $\delta$ perpendicular to $\gamma$. Then, we can expand the polarization to linear order in magnetic field to take into account the presence of a uniform, spatially independent magnetic field ~\cite{PhysRevB.90.125132} 
%
%
%
\begin{align}
 \lim_{\mathbf{A',q'\rightarrow 0}}&\left(\partial_{q'_\gamma} \partial_{A'_\delta}\Pi_{\alpha \beta}({\bf q}',\nu_n)\right){q'_\gamma} {A'_\delta}=\\&\lim_{\mathbf{A',q'\rightarrow 0}}\left(\partial_{q'_\gamma} \partial_{A'_\delta}\Pi_{\alpha \beta}({\bf q}',\nu_n)\right)(-i B^\eta)
\end{align}
where we used ${\bf B}({\bf q}') = i{\bf q}'\times {\bf A}'({\bf q}')$. Here, ${\bf q}'$ denotes the wave vector of the magnetic field that must be taken equal to zero along with ${\bf A}'$ at the end of the calculation, as indicated. As mentioned earlier, the total gauge potential is ${\bf A}(t)+{\bf A}'({\bf r})$ so that the frequency dependent Hall conductivity $\sigma^{H(\eta)}_{\alpha\beta}$ is given by  
\begin{align}\label{eq:Hall_conductivity}
	\sigma^{H(\eta)}_{\alpha\beta}&(i\nu_n)E^\beta(i\nu_n)=\frac{\Pi^{H(\eta)}_{\alpha \beta}(\nu_n)}{i\nu_n} B E^\beta(i\nu_n)\equiv \nonumber \\
	&\lim_{\mathbf{A',q'\rightarrow 0}}\left(-i\frac{\epsilon^{\eta\gamma\delta}}{2}\partial_{q'_\gamma} \partial_{A'_\delta}\frac{\Pi_{\alpha \beta}({\bf q}',\nu_n)}{i\nu_n}\right)B E^\beta(i\nu_n) 
\end{align}
where $B$ is the magnitude of the magnetic field, which points in the direction $\eta$ perpendicular to $\alpha$ and $\beta$. To write the result in a gauge invariant form, we used the Levi-Civita tensor. 
The first term (diamagnetic term) in the polarization tensor \eref{Pi1}, does not contribute to the Hall conductivity, anomalous or normal, because of the Kronecker $\delta_{\alpha\beta}$.  

The calculation of the derivatives with respect to ${\bf A}'$ and to ${\bf q}'$ is illustrated schematically in Fig.~\ref{fig:Diag}. A momentum $\mathbf{q}'$ comes in through the initial current vertex and comes out through a vertex associated with ${\bf A}'$. On the first line, the first diagram on the right-hand side comes from the derivative of the inital current vertex with respect to ${\bf A}'$ while, on the second line, the first and last diagrams come from using the matrix identity \eref{eq:dG=-GdG-1G} for the derivative of a Green function. The middle diagram comes from the derivative of the dressed current vertex. Computation of this derivative is rather subtle, as discussed in Appendix~\ref{Appendix_A}. It leads to the dressing of the bare current vertex $\lambda_\alpha$ which becomes $\Lambda_\alpha$. Using the same matrix identity \eref{eq:dG=-GdG-1G} for the derivative with respect to $\mathbf{q}'$, it is clear that four-leg (rectangular) diagrams will appear. The algebraic details are shown in Appendix~\ref{D}. 

Dividing the diagrams that enter $\Pi^{(\eta)}_{\alpha \beta}(\nu_n)$ into triangular and rectangular diagrams, we write
\begin{equation}
\Pi^{H(\eta)}_{\alpha \beta}(\nu_n)\simeq \Pi^{H(\eta), \rm{tri}}_{\alpha \beta}(\nu_n)+\Pi^{H(\eta), \rm{rec}}_{\alpha \beta}(\nu_n),\label{eq:Hall}
\end{equation}
where $\eta$ denotes the direction of the field, and we have {\it neglected diagrams that come from the three-particle scattering vertices} (see Appendix~\ref{Appendix_A} for more detail and Ref.~\cite{PhysRevB.94.045107}).  The three-leg, triangular, contribution comes from derivatives with respect to ${\bm q}_\delta'$ of the $\Lambda_\beta$ vertex in the first and last diagram of the second line in \fref{fig:Diag} and from derivatives with respect of ${\bm q}_\delta'$ of the Green's function of the middle diagram. The terms combine to give
\begin{widetext}
\begin{align}
\Pi^{H(\eta), \rm{tri}}_{\alpha \beta}(\nu_n)= (\frac{-i\epsilon^{\eta\delta\gamma}Bk_BT}{2V_{\rm cell}N})\frac{\hbar}{e}\sum_{{\bf k},\omega_m}&{\rm Tr}\bigg\{
\bigg( \left[{\bm \Lambda}_{\alpha}(\nu_n)\right]^{{\bf k}}_{\omega_m;\omega_m^+}
{\bf G}_{{\bf k}, \omega_m^+}
\left[{\bm \Lambda}_{\delta \beta}( -\nu_n)\right]^{{\bf k}}_{\omega_m^+;\omega_m}
{\bf G}_{{\bf k}, \omega_m}
\left[{\bm \Lambda}_{\gamma}( 0)\right]^{{\bf k}}_{\omega_m;\omega_m}
{\bf G}_{{\bf k}, \omega_m}
\bigg) \nonumber\\
-&\bigg(\left[{\bm \Lambda}_{\alpha}(\nu_n)\right]^{{\bf k}}_{\omega_m^-;\omega_m}{\bf G}_{{\bf k}, \omega_m}
\left[{\bm \Lambda}_{\gamma}( 0)\right]^{{\bf k}}_{\omega_m;\omega_m}
{\bf G}_{{\bf k}, \omega_m}\left[{\bm \Lambda}_{\delta \beta}( -\nu_n)\right]^{{\bf k}}_{\omega_m;\omega_m^-}
{\bf G}_{{\bf k}, \omega_m^-}
\bigg)\bigg\}, \label{eq:Tri}
\end{align}

\begin{figure}[t!]
	\begin{center}
		\begin{tabular}{c}
			\includegraphics[width=0.57\columnwidth]{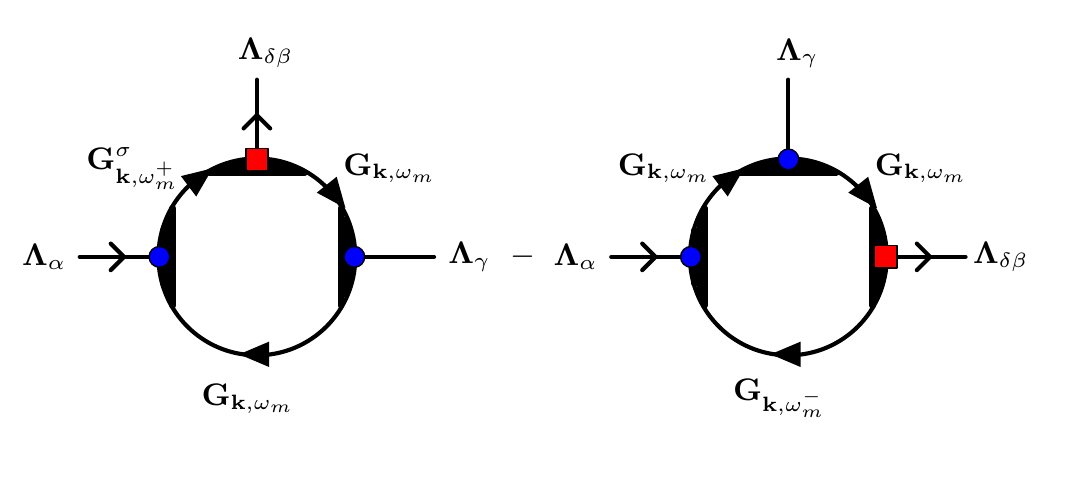}  
		\end{tabular} 
		\caption{The triangular diagrams appearing at the Hall response of multi-band systems.  The arrows on the vertices show the incoming or outcoming bosonic frequency associated with the electric field frequency. The circle and square vertices demonstrate current vertex and current vertex derivative (effective mass), repectively.}\label{fig:HallDiagTri}
	\end{center}
\end{figure}
\begin{figure}[t!]
	\begin{center}
		\begin{tabular}{c}
			\includegraphics[width=0.9\columnwidth]{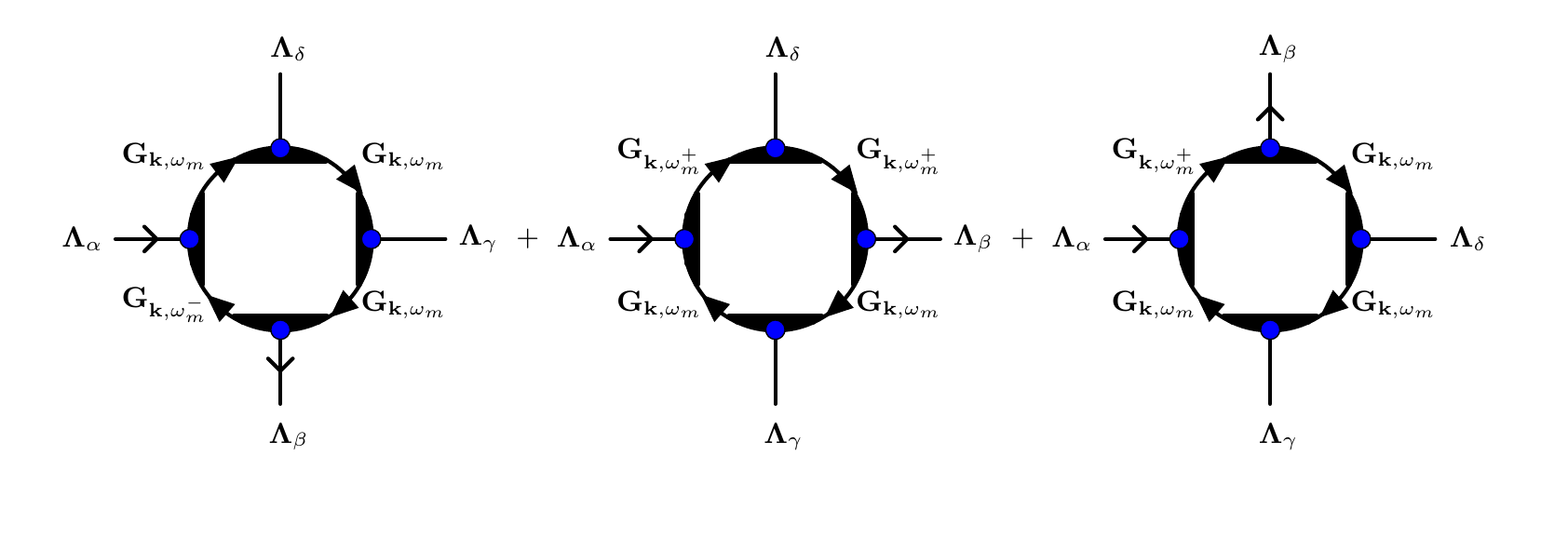} 
		\end{tabular} 
		\caption{The rectangular diagrams appearing at the Hall response of multi-band systems. These diagrams are absent in the single-band system.  The arrows on the vertices show the incoming or outcoming bosonic frequency associated with the electric field frequency. All the veritces are current vertices.}\label{fig:HallDiagRec}
	\end{center}
\end{figure}
\end{widetext}
where we used a compact notation for the dressed current vertices  
\begin{equation}
\left[{\bm \Lambda}_{\alpha}(\nu_n)\right]^{{\bf k}}_{\omega_m^-;\omega_m} \equiv 
\left[{\bm \Lambda}_{\alpha}({\bm 0}, \nu_n)\right]_{{\bf k},\omega_m^-;{\bf k},\omega_m},
\end{equation}
and defined $\omega_m^{-}\equiv \omega_m - \nu_n$. These dressed current vertices obey the usual integral equation \eref{eq:Integral_eq_for_dressed_vertex}. We also used the following notation for the dressed inverse effective mass tensor 
\begin{equation}
\left[{\bm \Lambda}_{\delta\beta}(\nu_n)\right]^{{\bf k}}_{\omega_m^+;\omega_m} \equiv 
\left[{\bm \Lambda}_{\delta\beta}({\bm 0}, \nu_n)\right]_{{\bf k},\omega_m^+;{\bf k},\omega_m},
\end{equation}
which obeys an integral equation similar to that of the dressed current vertex, as given in \eref{eq:def_dressed_effective_mass}.

The Feynman diagrams for the three-leg contribution are shown at \fref{fig:HallDiagTri}. 
The triangular contribution can be obtained from a direct extension of the single-band formula~\cite{PhysRevB.45.13945, Prog.Theor.Phys.80.623}.

The four-leg, rectangular, contribution is obtained from the derivative with respect to ${\bm q}_\delta'$ of the Green's functions in the first and last diagrams in the second line of \fref{fig:Diag}. The results combine to give 
\begin{widetext}
\begin{align}
\Pi^{H(\eta), \rm{rec}}_{\alpha \beta}(\nu_n)= &(\frac{-i\epsilon^{\eta\delta\gamma}Bk_BT}{2V_{\rm cell}N})\frac{\hbar}{e}\sum_{{\bf k},\omega_m}{\rm Tr}\bigg\{\nonumber \\
&\left[{\bm \Lambda}_{\alpha}(\nu_n)\right]^{{\bf k}}_{\omega_m^-;\omega_m}
{\bf G}_{{\bf k},\omega_m}
\left[{\bm \Lambda}_{\delta}(0)\right]^{{\bf k}}_{\omega_m;\omega_m}
{\bf G}_{{\bf k},\omega_m}
\left[{\bm \Lambda}_{\gamma}(0)\right]^{{\bf k}}_{\omega_m;\omega_m}
{\bf G}_{{\bf k}, \omega_m}
\left[{\bm \Lambda}_{\beta}(-\nu_n)\right]^{{\bf k}}_{\omega_m;\omega_m^-}
{\bf G}_{{\bf k}, \omega_m^-}\nonumber\\
+&\left[{\bm \Lambda}_{\alpha}(\nu_n)\right]^{{\bf k}}_{\omega_m;\omega_m^+}
{\bf G}_{{\bf k},\omega_m^+}
\left[{\bm \Lambda}_{\delta}(0)\right]^{{\bf k}}_{\omega_m^+;\omega_m^+}
{\bf G}_{{\bf k},\omega_m^+}
\left[{\bm \Lambda}_{\beta}(-\nu_n)\right]^{{\bf k}}_{\omega_m^+;\omega_m}
{\bf G}_{{\bf k}, \omega_m}
\left[{\bm \Lambda}_{\gamma}(0)\right]^{{\bf k}}_{\omega_m;\omega_m}
{\bf G}_{{\bf k}, \omega_m}\nonumber\\
+&\left[{\bm \Lambda}_{\alpha}(\nu_n)\right]^{{\bf k}}_{\omega_m;\omega_m^+}
{\bf G}_{{\bf k},\omega_m^+}
\left[{\bm \Lambda}_{\beta}(-\nu_n)\right]^{{\bf k}}_{\omega_m^+;\omega_m}
{\bf G}_{{\bf k}, \omega_m}
\left[{\bm \Lambda}_{\delta}(0)\right]^{{\bf k}}_{\omega_m;\omega_m}
{\bf G}_{{\bf k},\omega_m}
\left[{\bm \Lambda}_{\gamma}(0)\right]^{{\bf k}}_{\omega_m;\omega_m}
{\bf G}_{{\bf k}, \omega_m}
\bigg\},\label{eq:Rec}
\end{align} 
\end{widetext} 
and are shown in \fref{fig:HallDiagRec}. 

In a single band system, the rectangular part sums up to zero: Indeed, in that case all the matrices become scalars so they commute. Hence, the first and last line both vanish because they are symmetric in the indices $\delta$ and $\gamma$ and they are multiplied by $\epsilon^{\eta\delta\gamma}$. 
The middle line is also symmetric in $\delta$ and $\gamma$ if $\left[{\bm \Lambda}_{\delta}(0)\right]^{{\bf k}}_{\omega_m^+;\omega_m^+}=\left[{\bm \Lambda}_{\delta}(0)\right]^{{\bf k}}_{\omega_m;\omega_m}$ which we expect to be the case for most vertex corrections. For a multi-band system, however, some interband contributions survive and should be taken into account, as we show explicitly for the non-interacting limit in the next section. 

To compute the Hall conductivity, one adds, as in \eref{eq:Hall}, the three-leg \eref{eq:Tri} and the four-leg \eref{eq:Rec} results in Matsubara frequency, substitutes them in the expression for the Hall conductivity in terms of polarization \eqref{eq:Hall_conductivity} and performs analytic continuation.

From \eref{eq:Hall}, \eref{eq:Tri} and \eref{eq:Rec} one can see that $\Pi^{H(\eta)}_{\alpha \beta}(\nu)$ is an odd 
function of the bosonic Matsubara frequency, while the longitudinal current-current susceptibility is an even function. Therefore, the Hall susceptibility vanishes in zero bosonic Matsubara frequency. The Hall conductivity is also zero in the particle-hole (p-h) symmetric system. Again, this can be easily seen in the absence of the vertex correction from \eref{eq:Tri} and \eref{eq:Rec} where vertices are frequency independent. Due to p-h symmetry the Green’s function satisfy the identity ${\bf G}_{{\bf k}, \omega_m}={\bf G}_{-{\bf k}, -\omega_m}$, which, along with symmetry property of the vertices, leads to a vanishing Hall conductivity. In other word, electrons and holes drifting in perpendicular magnetic and electric fields (or in a temperature gradient) give contributions to the electric current of opposite sign. Therefore, Hall (and thermoelectric) currents arise due to the difference between electron and hole state, i.e., due to p-h asymmetry.

\vspace{12pt}

\section{DC limit and semiclassical transport theory}\label{Sec:DC_Semiclassical}
Here we show that the equations that we wrote for the longitudinal and Hall conductivities are basis independent and  reduce to the semiclassical theory at the DC limit. The semiclassical theory neglects the real and virtual interband processes in which an electron may change its band index as it traverses the Brillouin zone. Hence, the current vertices and their derivatives are diagonal matrices with the diagonal elements equal to the first and  the second derivative of the band energies, respectively. By contrast, in the orbital basis, interband transitions are allowed. 


Our approach will be, for the non-interacting case, to write the formulas for the DC longitudinal and Hall conductivities in the orbital basis and to show that the results are identical to those of the semi-classical theory, which is written in the band basis. 

Assume that the Hamiltonian ${\bf H}_{{\bf k}}$ in the orbital basis is diagonalized by the unitary matrix ${\bf U}_{{\bf k}}$ as follows: ${\bf U}_{{\bf k}}{\bf H}_{{\bf k}}{\bf U}^{\dagger}_{{\bf k}}$.  In the limit of ${\bf q} \rightarrow {\bm 0}$ and ${\bf A} \rightarrow {\bm 0}$ one can replace $\partial_{A_{\alpha}}$ by $(e/\hbar) \partial_{k_{\alpha}}$, where $e$ denotes elementary charge. 
Thus,  ${\bm \lambda}_{\sigma}^{\alpha}$ can be substituted by $ -(e/\hbar)\partial_{k_\alpha}{\bf H}_{0} $ in this limit (minimal coupling limit). This approximation neglects intra-atomic excitations. For non-interacting systems the current vertices are bare vertices and  in the orbital basis they are given by ${\bm \lambda}_{{\bf k}}^{\alpha}=-(e/\hbar)\partial_{k_\alpha}{\bf H}$ and ${\bm \lambda}_{{\bf k}}^{\alpha\beta}=(e/\hbar)^2\partial_{k_\alpha}\partial_{k_\beta}{\bf H}$. 

When we perform a change of basis in the trace expressions for conductivities, the vertices become  $\tilde{{\bm \lambda}}_{{\bf k}}^{\alpha}\equiv {\bf U}_{{\bf k}}{\bm \lambda}_{{\bf k}}^{\alpha}{\bf U}^{\dagger}_{{\bf k}}$. They have both intra- and inter-band components. The proofs below, rest on the following identities~\cite{doi:10.7566/JPSJ.84.124708} 
\begin{align}
-(e/\hbar)\partial_{k_{\alpha}}E_{n,{\bf k}}&=\big[\tilde{{\bm \lambda}}_{{\bf k}}^{\alpha}\big]_{nn}, \label{Jintra-band} \\
(e/\hbar)^2\partial_{k_{\alpha}}\partial_{k_{\beta}}E_{n,{\bf k}}&=\big[\tilde{{\bm \lambda}}_{{\bf k}}^{\alpha\beta}\big]_{nn}\nonumber\\+ 
\sum_{n'(\neq l)}&\frac{\left[\tilde{{\bm \lambda}}_{{\bf k}}^{\alpha}\right]_{nn'}\left[\tilde{{\bm \lambda}}_{{\bf k}}^{\beta}\right]_{n'n}+\left[\tilde{{\bm \lambda}}_{{\bf k}}^{\beta}\right]_{nn'}\left[\tilde{{\bm \lambda}}_{{\bf k}}^{\alpha}\right]_{n'n}}{E_{n,{\bf k}}-E_{n',{\bf k}}}\label{dJintra-band},
\end{align}
where $n$ in $E_{n,{\bf k}}$ is the band index.  Note that the last term in \eref{dJintra-band} is large only if the band energies are close to each other.

\subsection{Longitudinal semi-classical conductivity}
We start with the longitudinal response, \eref{Pi1}. The intra-band contribution of the paramagnetic current is equal to the semiclassical contribution because diagonal elements of the current operator are identical in both basis, as can be seen in \eref{Jintra-band}. However, the paramagnetic current has  inter-band contributions as well that, in DC limit, are given by
\begin{widetext}
\begin{align}
\Pi^{interband}_{\alpha \alpha}({\bm 0},\nu_n  \rightarrow 0)&=
\frac{-k_BT}{2V_{\rm cell}N}
\frac{\hbar}{e}
\sum_{{\bf k},\omega_m}\sum_{n,n'(\neq n)} 
\left[\tilde{{\bm \lambda}}_{{\bf k}}^{\alpha}\right]_{nn'}
\left[{\bf G}_{{\bf k},\omega_m^+}\right]_{n'n'}
\left[\tilde{{\bm \lambda}}_{{\bf k}}^{\alpha}\right]_{n'n}
\left[{\bf G}_{{\bf k},\omega_m}\right]_{nn}\nonumber\\
&=\frac{-1}{2V_{\rm cell}N}\frac{\hbar}{e}\sum_{{\bf k}}\sum_{n,n'(\neq n)} 
\frac{2\left[\tilde{{\bm \lambda}}_{{\bf k}}^{\alpha}\right]_{nn'}
\left[\tilde{{\bm \lambda}}_{{\bf k}}^{\alpha}\right]_{n'n}}{E_{n,{\bf k}}-E_{n',{\bf k}}}f(E_{n,{\bf k}})\label{eq:J1}
\end{align}
\end{widetext}
where the non-interacting Green's function in the band basis is diagonal and 
is given by
\begin{equation}\label{eq:GreenGamma}
\left[{\bf G}_{{\bf k},\omega_m}\right]_{nn} =\left[ i\omega_m+\mu - E_{n,{\bf k}} +i{\rm sgn}(\omega_m)\Gamma\right]^{-1},
\end{equation}
where $1/\tau=2\Gamma$ denotes the scattering rate from impurities. We neglect impurity vertex corrections and weak localization corrections. We also assume $\Gamma$ is smaller than the energy gap. The electron spectral function is a narrow Lorentzian function.  $f(E_{n,{\bf k}})$ denotes the Fermi-Dirac distribution function. 
Since we are looking at $\nu_n \rightarrow 0$ limit,  the scattering rate is irrelevant and is omitted in Matsubara frequency summations in \eref{eq:J1} and \eref{eq:J2}.

On the other hand the diamagnetic term is given by
%
%
\begin{align}
\Pi^{dia}_{\alpha \alpha}({\bm 0},\nu_n  \rightarrow 0)=&\frac{-k_BT}{2V_{\rm cell}N}\frac{\hbar}{e}\sum_{{\bf k},\omega_m}\sum_n \bigg\{  \left[\tilde{{\bm \lambda}}_{{\bf k}}^{\alpha\alpha}\right]_{nn}
\left[{\bf G}_{{\bf k},\omega_m}\right]_{nn}\bigg\}  \nonumber \\
=&\frac{-1}{2V_{\rm cell}N}\frac{\hbar}{e}\sum_{{\bf k}}\sum_n \left[\tilde{{\bm \lambda}}_{{\bf k}}^{\alpha\alpha}\right]_{nn} f(E_{n,{\bf k}}).\label{eq:J2}
\end{align}
%
%
One can check that the sum of the interband contribution to the paramagnetic current \eref{eq:J1} and the diamagnetic term \eref{eq:J2} yields the same result as the diamagnetic term in the semiclassical theory. Indeed, in that case, the diamagnetic term is evaluated with $\partial_{k_{\alpha}}\partial_{k_{\alpha}}E_{n,{\bf k}}$ for the current vertex derivative (or inverse effective mass tensor) and the equality \eref{dJintra-band} for that quantity proves the result. Therefore, evaluating the longitudinal conductivity \eref{Pi1} in the DC limit using the orbital basis with the ${\bm \lambda}_{{\bf k}}^{\alpha}$ and ${\bm \lambda}_{{\bf k}}^{\alpha\beta}$ vertices, or the band basis, using the in the latter case the left hand side of \eref{Jintra-band} and  \eref{dJintra-band} for vertices, yield identical results.

Note that the inter-band terms contribute to the {\it transverse} conductivity to linear oder in $\nu_n$ 
\begin{align}
\Pi^{interband}_{\alpha \beta}&({\bm 0},\nu_n) \simeq\frac{-i\nu_n}{2V_{\rm cell}N}\frac{\hbar}{e}\sum_{{\bf k}}\nonumber\\
\times \sum_{n,n'(\neq n)} 
&\left[\tilde{{\bm \lambda}}_{{\bf k}}^{\alpha}\right]_{nn'}
\left[\tilde{{\bm \lambda}}_{{\bf k}}^{\beta}\right]_{n'n}\frac{f(E_{l,{\bf k}})-f(E_{n',{\bf k}})}{(E_{n,{\bf k}}-E_{n',{\bf k}})^2},\label{eq:J2-1}
\end{align}
which gives the Berry curvature contribution to the anomalous Hall conductivity. This contribution is dissipation-less~\cite{PhysRevB.90.125132, PhysRevB.88.155121} and  does not appear in the semiclassical theory, which only accounts for the dissipative part of the response.  The anomalous Hall effect (AHE) occurs in solids with broken time-reversal symmetry, typically in a ferromagnetic phase, or, as the above formula shows, in solids with non-trivial topology as a consequence of spin-orbit coupling~\cite{RevModPhys.82.1539}. The Berry curvature takes into account the admixture of occupied and unoccupied bands and becomes very small for large band gaps. In ferromagnetic systems the AHE is related to the sample spontaneous magnetization. 


\subsection{Semi-classical Hall conductivity}
A direct comparison with the semi-classical theory in Matsubara frequency cannot be done because there are finite frequency interband transitions that cannot be described by the semi-classical theory. The two approaches coincide only in the DC limit, which requires analytic continuation. Nevertheless, we can convince ourselves of the equivalence of the approaches by focusing on the lowest two Matsubara frequencies. 

The zero Matsubara-frequency result is trivial because the Hall response is an odd function of the bosonic Matsubara frequency, hence it vanishes in the $\nu_n \rightarrow 0$ limit. The leading term in the low temperature limit is $\nu_n \rightarrow \nu_1=2\pi k_BT$ limit. 

Starting with the triangular diagrams, the terms appearing in  $\Pi^{(\eta), \rm{tri}}_{\alpha \beta}(\nu_n)$ can be classified in terms of the band indices of the three Green's functions. Performing the fermionic Matsubara frequency summation, (Appendix \ref{Sec:Matsubara}) one can show that only intraband terms have non-zero contribution in the $\nu_n \rightarrow \nu_1$ limit (See Appendix \ref{Sec:Appendix_DC} for a discussion of interband terms).  
Therefore, $\Pi^{(\eta), \rm{tri}}_{\alpha \beta}(\nu_n\rightarrow \nu_1)$ is 
%
\begin{align}
\Pi&^{H(\eta), \rm{tri}}_{\alpha \beta}(\nu_n\rightarrow \nu_1)= (\frac{3\tau^2}{2\pi^2})(\frac{-i\epsilon^{\eta\delta\gamma}B}{2V_{\rm cell}N})\frac{\hbar}{e}\nonumber \\&\times \sum_{{\bf k},n}
\left[\tilde{{\bm \lambda}}_{{\bf k}}^{\alpha}\right]_{nn}
\left[\tilde{{\bm \lambda}}_{{\bf k}}^{\delta \beta}\right]_{nn}
\left[\tilde{{\bm \lambda}}_{{\bf k}}^{\gamma}\right]_{nn}
(\frac{2}{i\nu_n})\frac{\partial f(E_{n,{\bf k}})}{\partial E_{n,{\bf k}}}, \label{Tri-band}
\end{align}
%
where we have used \eref{eq:1} with $n_1=n_2$ and the spectral representation of the Green's function in the \emph{band basis}
\begin{equation}
{\bf G}({\bf k},i\omega_m) = \int d\omega \frac{{\bf \mathcal{A}}_{{\bf k}}(\omega)}{(i\omega_m-\omega)},
\end{equation}
where ${\bf \mathcal{A}}_{{\bf k}}(\omega)=(-\frac{1}{\pi})Im {\bf G}({\bf k},\omega)$. We also replaced the product of spectral functions with a delta function~\cite{PhysRevB.45.13945} 
\begin{equation}
[\mathcal{A}_{n{\bf k}}(\omega)]^3\rightarrow (3\tau^2/2\pi^2)\delta(\omega-E_{n,{\bf k}}),\label{eq:A}
\end{equation}
for small $\Gamma$. The relaxation time, $\tau$, is defined as $1/\tau=2\Gamma$.   

It is clear that if we just transform the above result to the band basis, $\Pi^{(\eta), \rm{tri}}_{\alpha \beta}(i\nu_n\rightarrow \nu_1)$ gives a different results. Indeed, in this case $\Pi^{(\eta), \rm{tri}}_{\alpha \beta}(\nu_n\rightarrow \nu_1)$ is given by a similar equation, but the curent vertices are replaced with first and second derivatives of the band energies. Since, $\left[\tilde{{\bm \lambda}}_{{\bf k}}^{\delta \beta}\right]_{nn}$ is different from the second derivative of the band energy, as seen in \eref{dJintra-band}, the result is basis dependent. The second part of the Hall response, given by the four-leg, rectangular, diagrams $\Pi^{(\eta), \rm{rec}}_{\alpha \beta}(\nu_n)$, when added to the triangular diagrams will be basis independent. 
The rectangular diagrams contain several classes of diagrams, but many of them vanish for $\nu_n \rightarrow \nu_1$. It is not complicated to show that the purely intraband terms in $\Pi^{(\eta), \rm{rec}}_{\alpha \beta}(\nu_n)$ cancel out. It is also straightforward but lengthy to show that in the $\nu_n \rightarrow \nu_1$ limit, the only diagrams with non-zero contributions are those where $\alpha, \gamma$ vertices do not change the band index while $\delta, \beta$ vertices change the band index (along with those resulting from $\gamma \leftrightarrow \delta$) (see Appendix \ref{Sec:Matsubara}). Furthermore $\alpha, \gamma$ vertices must have the same band index. To be explicit, consider terms with $\tilde{{\bm \lambda}}_{{\bf k},nn}^{\alpha}\tilde{{\bm \lambda}}_{{\bf k},nn}^{\gamma}$ and $\tilde{{\bm \lambda}}_{{\bf k},nn'}^{\beta}\tilde{{\bm \lambda}}_{{\bf k},n'n}^{\delta}$ or $\tilde{{\bm \lambda}}_{{\bf k},nn'}^{\delta}\tilde{{\bm \lambda}}_{{\bf k},n'n}^{\beta}$ with $n\neq n'$. The surviving terms come from the second and third lines on the right-hand side of \eref{eq:Rec} and they are given by
\begin{widetext}
\begin{align}
{\rm lim}_{\nu_n \rightarrow \nu_1}&(\frac{-i\epsilon^{\eta\delta\gamma}Bk_BT}{2V_{\rm cell}N})\frac{\hbar}{e}\sum_{{\bf k},\omega_m}\sum_{n,n'(\neq n)}\left[\tilde{{\bm \lambda}}_{{\bf k}}^{\alpha}\right]_{nn}\left[\tilde{{\bm \lambda}}_{{\bf k}}^{\gamma}\right]_{nn}
\left[{\bf G}_{{\bf k},\omega_m}\right]_{nn}\left[{\bf G}_{{\bf k},\omega_m}\right]_{nn}\bigg\{\nonumber \\
&\left( \left[\tilde{{\bm \lambda}}_{{\bf k}}^{\beta}\right]_{nn'}\left[\tilde{{\bm \lambda}}_{{\bf k}}^{\delta}\right]_{n'n}
\left[{\bf G}_{{\bf k},\omega_m}\right]_{n'n'}\left[{\bf G}_{{\bf k},\omega_m^+}\right]_{nn} 
+\left[\tilde{{\bm \lambda}}_{{\bf k}}^{\delta}\right]_{nn'}\left[\tilde{{\bm \lambda}}_{{\bf k}}^{\beta}\right]_{n'n}
\left[{\bf G}_{{\bf k},\omega_m^+}\right]_{nn}\left[{\bf G}_{{\bf k},\omega_m^+}\right]_{n'n'} 
\right) \bigg\}
\nonumber\\
=&(\frac{3\tau^2}{2\pi^2})(\frac{-i\epsilon^{\eta\delta\gamma}B}{2V_{\rm cell}N})\frac{\hbar}{e}\sum_{{\bf k}}\sum_{n,n'(\neq n)}\left[\tilde{{\bm \lambda}}_{{\bf k}}^{\alpha}\right]_{nn}\left[\tilde{{\bm \lambda}}_{{\bf k}}^{\gamma}\right]_{nn}
\frac{\left[\tilde{{\bm \lambda}}_{{\bf k}}^{\delta}\right]_{nn'}\left[\tilde{{\bm \lambda}}_{{\bf k}}^{\beta}\right]_{n'n}+\left[\tilde{{\bm \lambda}}_{{\bf k}}^{\delta}\right]_{n'n}\left[\tilde{{\bm \lambda}}_{{\bf k}}^{\beta}\right]_{nn'}}{E_{n,{\bf k}}-E_{n',{\bf k}}}
(\frac{1}{i\nu_n})\frac{\partial f(E_{n,{\bf k}})}{\partial E_{n,{\bf k}}},\label{Rec-band}
\end{align}
\end{widetext}
where we have used \eref{eq:3} and \eref{eq:4} and expanded them in terms of $\nu_n$, keeping non-vanishing terms in the $\nu_n \rightarrow \nu_1$ and low temperature limit. We also assumed that the chemical potential lies within band $l$ and replaced $i\omega_m$ in the band $l^\prime$ Green's function with the pole at $E_{n,{\bf k}}-\mu$. Hence the Green's function for this band is replaced by $1/(E_{n,{\bf k}}-E_{n',{\bf k}})$.~\cite{PhysRevB.45.13945} This allows us to employ \eref{eq:A}.

Using \eref{eq:3} and \eref{eq:5}, the terms with $\tilde{{\bm \lambda}}_{{\bf k},nn}^{\alpha}\tilde{{\bm \lambda}}_{{\bf k},nn}^{\delta}$ and $\tilde{{\bm \lambda}}_{{\bf k},nn'}^{\beta}\tilde{{\bm \lambda}}_{{\bf k},n'n}^{\gamma}$ or $\tilde{{\bm \lambda}}_{{\bf k},nn'}^{\gamma}\tilde{{\bm \lambda}}_{{\bf k},n'n}^{\beta}$ with $n\neq n'$ yield a result similar to the last line of \eref{Rec-band} but with a negative sign and $\gamma \leftrightarrow \delta$. Similar diagrams survive after operating with the Levi-Civita tensor. So finally, one needs to multiply \eref{Rec-band} by a factor two. 

The combination of \eref{Tri-band} with twice \eref{Rec-band} gives the semiclassical transport theory formula for the Hall effect. 

But, if one consider terms with $\tilde{{\bm \lambda}}_{{\bf k},nn}^{\beta}\tilde{{\bm \lambda}}_{{\bf k},nn}^{\gamma}$ vertices and $\tilde{{\bm \lambda}}_{{\bf k},nn'}^{\alpha}\tilde{{\bm \lambda}}_{{\bf k},n'n}^{\delta}$ or $\tilde{{\bm \lambda}}_{{\bf k},nn'}^{\delta}\tilde{{\bm \lambda}}_{{\bf k},n'n}^{\alpha}$, then the result from the first line of \eref{eq:Rec} cancels out the result the from third line. The second line of \eref{eq:Rec} does not contribute to the trace for this specific choice of band indices. Note that $\tilde{{\bm \lambda}}$ is a symmetric matrix in band indices. 
%

Releasing some of the the semi-classical approximations leads to some modifications that are discussed in more detail in Ref.~\onlinecite{1807.10252} for a non-interacting two-band model with constant scattering rate. 

\section{Conclusions}\label{Sec:Conclusions}
In conclusion we derived an exact formula for the Hall response of multi-band systems. 
The formula contains two sets of diagrams: (i) triangular and (ii) rectangular diagrams. To compute the Hall conductivity, one adds, as in \eref{eq:Hall}, the three-leg \eref{eq:Tri} and the four-leg \eref{eq:Rec} results in Matsubara frequency, substitutes them in the expression for the Hall conductivity in terms of polarization \eqref{eq:Hall_conductivity} and performs analytic continuation. The vertices are given by \eref{eq:bare_vertex},\eqref{eq:dressed_vertex},\eqref{eq:bare_inv_effective_mass} and \eref{eq:def_dressed_effective_mass}. For the current vertex coupled to the magnetic field, we can use \eqref{eq:dk=dA} without approximations. Interchanging $\alpha$ and $\beta$ changes the sign of the Hall conductivity. Our results are also valid at finite frequency, if one takes into account the fact that the $\alpha$ and $\beta$ current vertices can contain atomic dipole transitions. At finite frequency, the off-diagonal conductivity contributes to Faraday rotation.   

Apart from the neglect of three-particle $\delta^2 \Sigma / \delta G \delta G$ and four-particle $\delta^3 \Sigma / \delta G \delta G \delta G$ scattering vertices, our formula is general and can be used for multi-band systems with arbitrary band topology, in the presence of spin-orbit coupling and of interactions, where the semiclassical theory is not applicable. The combination of triangular and rectangular diagrams reduces to the semiclassical transport theory in the DC limit with weak correlations. Appendix \ref{Sec:Matsubara} is useful to simplify the result in the case of non-interacting electrons. 

The triangular diagrams can be obtained from a direct extension of the single-band formula.  The rectangular diagrams are absent for  single-band systems but essential to obtain a basis-independent result for multi-band systems. Note however that when intra-atomic dipole transitions can be neglected, the current vertices can be obtained from derivatives with respect to wave vector. In that case, one must be in the orbital basis with a definition of Fourier transforms that includes actual positions of atoms or use the velocity operator of Ref.~\onlinecite{PhysRevB.80.085117}. Transformation to the band basis is done after the correct current vertices are calculated. 

The triangular diagrams \eref{eq:Tri} do not contain any topological response. They vanish in the absence of an external field whether or not the system shows a spontaneous magnetization, so the corresponding Hall conductivity has a linear dependence on the external magnetic field that extrapolates to zero at zero magnetic field. However, non-linear behavior is expected for ferromagnetic conductors where the Hall conductivity shows an initial sharp enhancement at low field due to the saturation of the magnetization of the sample under external field~\cite{PhysRev.36.1503}. The linear dependence is recovered only at high field for ferromagnetic conductors. The initial enhancement of the Hall response is accounted for by the linear response of the anomalous Hall effect to the magnetic field. The latter response is given by the rectangular diagrams. 

A direct extension of the single band formula for the Nernst effect or thermal Hall effect would not give the correct formula for multi-band systems. Modifications similar to those discussed here are necessary.

\begin{acknowledgments}
We are grateful to M.~Charlebois and A. Georges for discussions that led to this work. M.~Charlebois performed numerical calculations for Appendix~\ref{app:AFM} that helped identify the importance of rectangular diagrams. We are also thankful to S.~Verret, O.~Simard and M.~Zingl for useful discussion and to  J.~Mitscherling and W.~Metzner for helpful correspondance. R.N. acknowledges J.~Tomczak for useful discussions at early stages of this work. This work has been supported by the Natural Sciences and Engineering Research Council of Canada (NSERC) under grant RGPIN-2014-04584, by the Canada First Research Excellence Fund, and by the Research Chair in the Theory of Quantum Materials. 
\end{acknowledgments}

\appendix

\section{Current and inverse effective mass tensor in model Hamiltonians: the antiferromagnet}\label{app:AFM}

We first recall the semi-classical formulae for the DC and Hall conductivities and then illustrate the ambiguities that arise with these formulae in the simple case of the commensurate antiferromagnet.

\subsection{Semiclassical DC and Hall conductivities}
Let us consider the case where the self-energy does not depend on momentum, as in dynamical mean-field theory for example. When the single-particle spectral weight $\mathcal{A}_{n,\mathbf k}(\omega)$ (with $\int d\omega \mathcal{A}_{n,\mathbf k}(\omega)=1$) is diagonal in band index and $E_{n,\mathbf{k}}$ is even under parity so that vertex corrections can be neglected, the usual linear response arguments, generalized naively to the multi-band case, lead to the following expression for the longitudinal DC conductivity  
\begin{align} \label{Eq:LongitudinalDC}
\sigma_{xx}=\frac{2\pi e^2}{V\hbar}\int d\omega \sum_{{\bf k}n}
\mathcal{A}^2_{n,\mathbf k}(\omega)
\Big(
\frac{\partial E_{n,\mathbf{k}}}{\partial k_x}\Big)^2
\left(-\frac{\partial f(\omega)}{\partial \omega}\right).
\end{align}

For the DC Hall conductivity, the corresponding frequently used expression is~\cite{PhysRevB.45.13945,storey_hall_2016,eberlein_fermi_2016-1,Chatterjee_Sachdev_2016,verret_phenomenological_2017,Morice-Pepin:2017,Charlebois_Verret:2017}:
\begin{align} \label{Eq:DC_Hall_simple}
\sigma_{xy}=\frac{2 \pi^2 e^3}{3V\hbar}B\int d\omega \sum_{{\bf k}n}
\mathcal{A}^3_{n,\mathbf k}&(\omega)\Big[
2 
\frac{\partial E_{n,\mathbf{k}}}{\partial k_x}
\frac{\partial E_{n,\mathbf{k}}}{\partial k_x}
\frac{\partial^2 E_{n,\mathbf{k}}}{\partial k_x \partial k_y}\nonumber
\\
-\Big(
\frac{\partial E_{n,\mathbf{k}}}{\partial k_x}\Big)^2
\frac{\partial^2 E_{n,\mathbf{k}}}{\partial k_y^2}- 
&\Big(
\frac{\partial E_{n,\mathbf{k}}}{\partial k_y}\Big)^2
\frac{\partial^2 E_{n,\mathbf{k}}}{\partial k_x^2}
\Big] \frac{\partial f(\omega)}{\partial \omega}.
\end{align}
with $B$ the magnetic field~\cite{Note1}.

\subsection{Antiferromagnet as an example}\label{Sec:AFM}

When there is more than one atom per unit cell, the current and effective masses differ depending on whether they are computed from a wave-vector derivative of the Hamiltonian in the orbital basis or in the band basis.~\cite{PhysRevB.80.085117} This is obvious since the unitary transformation between the Hamiltonian in the orbital and in the band basis depends on ${\mathbf k}$. Using the antiferromagnet as an example, we discuss the consequences, arguing for the correct choice. 

Assume that the Fermi level crosses a single band and that residual Hubbard-like interactions lead to an antiferromagnetic state. All other bands are either completely full or empty and they do not come in the discussion. Take as lattice vectors ${\bf a}_1=(1,0)a$ and ${\bf a}_2=(0,1)a$ and ion positions on the $A$ sublattice  ${\bf r}_A=(0,0)a$ and on the $B$ sublattice ${\bf r}_B=(1/2,1/2)a$. With first-neighbor hopping $t$ and second-neighbor hopping $t^\prime$, the mean-field Hamiltonian matrix in the orbital basis $(c^A_\mathbf{k},\,c^B_\mathbf{k})$ can be written, for one of the spin species, in the form
%
\begin{equation}
{\bf H}_{\bf k}=
\begin{bmatrix}  
\epsilon_d+\Delta+\zeta_\mathbf{k} & \xi_\mathbf{k} \\ 
\xi_\mathbf{k} & \epsilon_d-\Delta+\zeta_\mathbf{k}
\end{bmatrix}, \label{Ham1}
\end{equation}
%
with the definitions
\begin{align}  
\xi_\mathbf{k}&=-2t[\cos({\bf k}\cdot{\bm \delta}_1 )+\cos({\bf k}\cdot{\bm \delta}_2)]\\
\zeta_\mathbf{k}&=-2t'[\cos({\bf k}\cdot({\bm \delta}_1+ {\bm \delta}_2))+\cos({\bf k}\cdot({\bm \delta}_1- {\bm \delta}_2))].
\end{align}
where ${\bm \delta}_1=(1/2,1/2)a$ and ${\bm \delta}_2=(1/2,-1/2)a$ are nearst-neighbor bonds and $\epsilon_d$ is the on-site energy. 
%
A unitary transformation ${\bf U}_{\bf k}$ diagonalizes the Hamiltonian as follows ${\bf U}_{\bf k}{\bf H}_{\bf k}{\bf U}^{\dagger}_{\bf k}$. 

\paragraph{Current vertex}
Using the Peierls substitution, the current vertex in the $\alpha$ direction computed from a $\mathbf{k}$ derivative in the orbital basis is given by
\begin{equation}
{\bm \lambda}_{\alpha\bf k}=
-\partial_{A_\alpha}{\bf H}^{({\bf A})}_{\mathbf{k}}|_{{\bf A}={\bm 0}}=
-\frac{e}{\hbar}\begin{bmatrix}  
\partial_{k_{\alpha}}\zeta_{\bf k} & \partial_{k_{\alpha}}\xi_{\bf k}\\ 
\partial_{k_{\alpha}}\xi_{\bf k} & \partial_{k_{\alpha}}\zeta_{\bf k}
\end{bmatrix}. \label{current2}
\end{equation}
Note that we were able to replace the derivative with respect to vector potential by a derivative with respect to ${\bf k}$ because neglected intra-atomic dipole transitions, did the Peierls substitution and included the phase factor $\exp[i{\mathbf{k} \cdot (\mathbf{R}_i}-{\mathbf{R}_j})]$ for hopping within the unit cell as well as between unit cells.  Otherwise, if the phase factor includes only the position of the unit cell, the current cannot be expressed as a derivative with respect to ${\bf k}$ only.~\cite{PhysRevB.80.085117} Multiplying ${\bf U}_{\bf k}{\bf \lambda}_{\alpha\bf k}{\bf U}^{\dagger}_{\bf k}$ transforms this current to the band basis. We find
\begin{align}
{\bf U}_{\bf k}{\bm \lambda}_{\alpha\bf k}{\bf U}^{\dagger}_{\bf k}&=-\frac{e}{\hbar}\partial_{k_{\alpha}}\zeta_{\bf k}{\bm 1}
-\frac{e}{\hbar}\begin{bmatrix}  
-\sin\theta_{\bf k} & \cos\theta_{\bf k}\\ 
\cos\theta_{\bf k}& \sin\theta_{\bf k}
\end{bmatrix}\partial_{k_{\alpha}}\xi_{\bf k} \label{currentband2}
\end{align}
where $\tan\theta_{\bf k}=(\xi_\mathbf{k}/\Delta)$.
Since eigenenergies $E_{\pm}({\bf k})$ are given by
\begin{align}
E_{{\pm},{\bf k}}
=(\epsilon_{d}+\zeta_{\bf k})\pm \sqrt{\xi^2_{\bf k}+\Delta^2},
\end{align}
we find that only the diagonal elements of the orbital-basis current \eref{current2} are equal to the currents calculated from $-(e/\hbar)\partial_{k_{\alpha}}E_{\pm,\bf k}$. Hence, with diagonal spectral weights, the longitudinal DC conductivity obtained from \eref{Eq:LongitudinalDC} should be the same, regardless of how we define current vertices. If there are off-diagonal components to the spectral weight, then the answer is unambiguous only in the DC limit and with delta-function spectral weights since in that case interband transitions are prohibited by energy and momentum conservation. The finite frequency (optical) conductivity on the other hand depends sensitively on which of the above two current vertices are used. As argued in the main text, the correct choice is the derivative with respect to $k$ of the Hamiltonian in the orbital basis.

\paragraph{Effective masses} The derivatives of the current vertex with respect to wave vector give a measure of the inverse effective masses that are needed for the calculation of the Hall effect~\cite{Note2}.
Taking derivatives with respect to $\mathbf{k}$ of the model Hamiltonian in the orbital basis, we find
\begin{align}
{\bm \lambda}_{\alpha\beta\bf k}=
\partial_{A_\alpha}\partial_{A_\beta}{\bf H}^{({\bf A})}_{\mathbf{k}}|_{{\bf A}={\bm 0}}
=(\frac{e}{\hbar})^2\begin{bmatrix}  
\partial_{k_{\alpha}}\partial_{k_{\beta}}\zeta_{\bf k} & \partial_{k_{\alpha}}\partial_{k_{\beta}}\xi_{\bf k}\\ 
\partial_{k_{\alpha}}\partial_{k_{\beta}}\xi_{\bf k} & \partial_{k_{\alpha}}\partial_{k_{\beta}}\zeta_{\bf k}
\end{bmatrix}. \label{effectivemass2}
\end{align}
Transforming this to the band basis using ${\bf U}_{\bf k}{\bf \lambda}_{\alpha\beta\bf k}{\bf U}^{\dagger}_{\bf k}$ we find
\begin{align}
{\bf U}_{\bf k}{\bm \lambda}_{\alpha\beta\bf k}{\bf U}^{\dagger}_{\bf k}&=(e/\hbar)^2
\partial_{k_{\alpha}}\partial_{k_{\beta}}\zeta_{\bf k}{\bm 1}\nonumber\\&+
(\frac{e}{\hbar})^2\begin{bmatrix}  
-\sin\theta_{\bf k} & \cos\theta_{\bf k}\\ 
\cos\theta_{\bf k}& \sin\theta_{\bf k}
\end{bmatrix}\partial_{k_{\alpha}}\partial_{k_{\beta}}\xi_{\bf k}. \label{EffectiveMassband2}
\end{align}
This time, even the diagonal elements are different from what we find directly from $(e/\hbar)^2\partial_{k_{\alpha}}\partial_{k_{\beta}}E_{\pm,\bf k}$, which now is given by
\begin{align}
\partial_{k_{\alpha}}\partial_{k_{\beta}}E_{\pm,\bf k}&=\partial_{k_{\alpha}}\partial_{k_{\beta}}\zeta_{\bf k}\pm\big[\sin\theta_{\bf k}\partial_{k_{\alpha}}\partial_{k_{\beta}}\xi_{\bf k}\nonumber\\
+(&\cos\theta_{\bf k}/\Delta)(1-\sin^2\theta_{\bf k})\partial_{k_{\alpha}}\xi_{\bf k}\partial_{k_{\beta}}\xi_{\bf k}\big].\label{mass2}
\end{align}
This means that even for the DC Hall conductivity, \eref{Eq:DC_Hall_simple}, the answer depends on which inverse effective-mass tensor we use (when $t^\prime$ does not vanish). With the three-leg formula \eref{Eq:DC_Hall_simple}, Ref.~\onlinecite{PhysRevB.45.13945} has shown, surprisingly, that 
it is the band-basis inverse effective-mass $\partial_{k_{\alpha}}\partial_{k_{\beta}}E_{\pm,\bf k}$ that must be used. This can be confirmed by an explicit calculation~\cite{Charlebois_unpublished} which shows that 
the Hall number of a hole-doped mean-field antiferromagnetic band calculated with the three-leg formula \eref{Eq:DC_Hall_simple} is correct only if the effective mass $\partial_{k_{\alpha}}\partial_{k_{\beta}}E_{\pm,\bf k}$ is used. 

The above result seems paradoxical but it does not contradict our earlier choice of current vertex. Indeed, the derivation of Voruganti {\it et al.}~\cite{PhysRevB.45.13945} shows implicitely that four-leg diagrams must be included to obtain the correct result with the physically correct orbital-basis currents and inverse effective masses. They considered however only the one-band case split by antiferromagnetic order in the DC limit. In what follows, we present the derivation of the most general formula for the longitudinal conductivity and Hall conductivity, valid for arbitrary interactions, spin-orbit coupling, topology and number of bands, with or without broken symmetry. 


\section{Equations of motions for the derivative of the dressed-current vertex and dressing of the vertex of the measured current}\label{Appendix_A}
In this section we derive the equation of motion for the derivative of the dressed current vertex. The derivative of the bare current vertex is the inverse of the effective mass tensor, but the situation is more subtle in the dressed case. We show that while the current vertex is governed by the standard Bethe-Salpeter equation, the derivative of the dressed current vertex has more structure. It obeys a different equation which leads to different types of vertex corrections.  

Consider the paramagnetic term of conductivity in presence of an external magnetic field. Here, we work in real space and define $1\equiv ({\bf r}_1,\tau_1)$. Then the paramagnetic contribution is proportional to
\begin{equation}
{\rm Tr}\left( {\bm \lambda}_{\alpha}^{ ({\bm A}')}(1'1){\bf G}^{ ({\bm A}')}(12){\bm \Lambda}_{\beta}^{ ({\bm A}')}(22') {\bf G}^{ ({\bm A}')}(2'1')\right),\label{E121}
\end{equation}
where the trace is over all internal degrees of freedom. 
To simplify the calculation, we use a matrix notation for the Green's function and for the vertices, dropping the spin and potential vector dependence. Then, with the trace assumed implicitly, the above equation reads
\begin{align}
\lambda_\alpha G (\partial_{A_\beta}G^{-1})G.
\end{align}
Using $\partial G=-G (\partial G^{-1})G$, and the definition already given in \eref{eq:bare_inv_effective_mass}
\begin{align}
\lambda_{\gamma\alpha}\equiv\partial_{A_\gamma}{\bm \lambda}_{\alpha}^{\sigma ({\bm A}')}=-\partial_{A_\gamma}\partial_{A_\beta}{\bf{H}} ^\sigma
\label{eq:def_lambda}
\end{align}
the derivative with respect to $A_{\gamma}$ of the paramagnetic term \eref{E121} is given by 
\begin{align}\label{eq:dA_paramagnetic}
&\lambda_{\gamma\alpha} G (\partial_{A_\beta}G^{-1})G \nonumber\\
-&\lambda_\alpha G (\partial_{A_\gamma}G^{-1})G (\partial_{A_\beta}G^{-1})G \nonumber\\
 -&\lambda_\alpha G (\partial_{A_\beta}G^{-1})G (\partial_{A_\gamma}G^{-1})G \nonumber\\
+&\lambda_\alpha G (\partial_{A_\gamma}\partial_{A_\beta}G^{-1})G.
\end{align}
The fourth line is the derivative of the dressed current vertex. 

Using the expression for $G^{-1}$ and the chain rule for the self-energy, the dressed vertex $\partial_{A_\beta}G^{-1}$ obeys the following integral equation
\begin{align}\label{eq:Integral_eq_for_dressed_vertex}
\partial_{A_\beta}G^{-1}=\lambda_\beta+\frac{\delta\Sigma}{\delta G}[G(\partial_{A_\beta}G^{-1})G].
\end{align}
The matrix multiplication convention applies to the Green's functions in square parenthesis in the above equation. Recalling our use of the chain rule, the matrix indices of the resulting product in square parenthesis are identical to the matrix indices of the Green's function in the denominator of ${\delta\Sigma}/{\delta G}$. The latter quantity is the four-point vertex that is irreducible in the longitudinal particle-hole channel, i.e., ${\delta\Sigma}/{\delta G}\equiv \Gamma^{ir}$.

The derivative $\partial_{A_\gamma}(\partial_{A_\beta}G^{-1})$ of the dressed current vertex $(\partial_{A_\beta}G^{-1})$ can be written as 
\begin{align}
\partial_{A_\gamma}\partial_{A_\beta}G^{-1}=\lambda_{\gamma\beta}-\partial_{A_\gamma}\partial_{A_\beta}\Sigma.  
\end{align}
With the chain rule for the derivative of the self-energy, this becomes
\begin{align}
\partial_{A_\gamma}\partial_{A_\beta}G^{-1}=\lambda_{\gamma\beta}-\partial_{A_\gamma}\Gamma^{ir}\left[\partial_{A_\beta} G\right].  
\end{align}
{\it We now assume that the 
three-particle scattering vertex $\delta^2 \Sigma / \delta G \delta G=\delta \Gamma^{ir}/\delta G$ can be neglected.} Then, the above equation becomes
\begin{align}
\partial_{A_\gamma}\partial_{A_\beta} G^{-1}&\simeq\lambda_{\gamma\beta}-\Gamma^{ir}\left[\partial_{A_\gamma}\partial_{A_\beta} G\right]\nonumber\\
&=\lambda_{\gamma\beta}+\Gamma^{ir}\partial_{A_\gamma}\left[G(\partial_{A_\beta} G^{-1})G\right]\nonumber\\
&=\lambda_{\gamma\beta}+\Gamma^{ir}\left[G(\partial_{A_\gamma}\partial_{A_\beta} G^{-1})G\right]\nonumber\\
&-\Gamma^{ir}\left[G(\partial_{A_\gamma} G^{-1})G(\partial_{A_\beta} G^{-1})G\right]\nonumber\\
&-\Gamma^{ir}\left[G(\partial_{A_\beta} G^{-1})G(\partial_{A_\gamma} G^{-1})G\right].
\end{align}
The above equation is an integral equation for $\partial_{A_\gamma}\partial_{A_\beta} G^{-1}$.

In the equation for the derivative of the paramagnetic term \eref{eq:dA_paramagnetic} we now use the cyclic property of the trace to move the rightmost $G$ to the left and then substitute the above equation for ${\partial^2 G^{-1}}/{\partial{A_\gamma}\partial{A_\beta}}$. We find
\begin{align}\label{eq:dA_paramagnetic_2}
&G\lambda_{\gamma\alpha} G (\partial_{A_\beta}G^{-1})
\nonumber \\
-&G\lambda_\alpha G (\partial_{A_\gamma}G^{-1})G (\partial_{A_\beta}G^{-1}) \nonumber\\
- &G\lambda_\alpha G \Gamma^{ir} \left[G(\partial_{A_\gamma}G^{-1})G (\partial_{A_\beta}G^{-1})G\right] \nonumber\\
 -&G\lambda_\alpha G (\partial_{A_\beta}G^{-1})G (\partial_{A_\gamma}G^{-1}) \nonumber\\ 
-&G\lambda_\alpha G \Gamma^{ir} \left[ G(\partial_{A_\beta}G^{-1})G (\partial_{A_\gamma}G^{-1})G\right] \nonumber\\ 
+&G\lambda_\alpha G \lambda_{\gamma\beta}  \nonumber\\ 
+&G\lambda_\alpha G \Gamma^{ir} \left[ G(\partial_{A_\gamma}\partial_{A_\beta}G^{-1})G\right]
\end{align}
Continuing the iteration, one can check that we are generating the infinite series for the vertex corrections of all the $\lambda_\alpha$ vertices and also of the $\lambda_{\alpha\beta}$ vertex. 

Using the notation already given in \eref{eq:dressed_vertex}
\begin{equation}
\Lambda_\beta=(\partial_{A_\beta}G^{-1})
\label{eq:def_Lambda}
\end{equation}
for the dressed vertex that obeys the integral equation \eref{eq:Integral_eq_for_dressed_vertex}, the final form of equation \eqref{eq:dA_paramagnetic_2} is 
\begin{align}\label{eq:dA_paramagnetic_3}
&G\lambda_{\gamma\alpha} G \Lambda_\beta \nonumber\\
-&G\Lambda_\alpha G \Lambda_\gamma G \Lambda_\beta \nonumber\\
 -&G\Lambda_\alpha  G \Lambda_\beta G \Lambda_\gamma \nonumber\\
+&\Lambda_\alpha  G \Lambda_{\gamma\beta} G,
\end{align}
with the dressed inverse effective-mass tensor $\Lambda_{\gamma\beta}$ defined by the following integral equation 
\begin{align}
\Lambda_{\gamma\beta}&=\lambda_{\gamma\beta}+\Gamma^{ir}[G\Lambda_{\gamma\beta}G]\\ 
&=\lambda_{\gamma\beta}+\frac{\delta\Sigma}{\delta G}[G\Lambda_{\gamma\beta}G]
\label{eq:def_dressed_effective_mass}
\end{align}
which is analogous to the corresponding \eref{eq:Integral_eq_for_dressed_vertex} for the dressed current vertex $\Lambda_\beta=(\partial_{A_\beta}G^{-1})$.

To dissipate any ambiguities that the above notation may contain, we now give an example with indices restored. The current vertex is given by
\begin{align}
{\bm \Lambda}_{\beta} (22') &= \frac{\partial {\bf G}^{-1}_{ }(22')}{\partial {\bf A}_{\beta}(33')}=-\frac{\partial {\bf H} (22')}{\partial {\bf A}_{\beta}(33')}\nonumber \\&+\frac{\partial {\bm \Sigma}(22')}{\partial {\bf G}^{' }(44')}
{\bf G}(45)
\frac{\partial {\bf G}^{-1}(55')}{\partial {\bf A}_{\beta}(33')}
{\bf G}(5'4').
\end{align}
We need to know how the dressed vertex varies upon applying an external magnetic field. In other words
\begin{widetext}
\begin{align}
\frac{\partial {\bm \Lambda}_{\beta}^{({\bf A}')}(22')}{\partial {\bf A}'_{\gamma}(66')}\bigg|_{{\bf A}'={\bm 0}}=& -\frac{\partial^2 {\bf H} ^{({\bf A}')}(22')}{\partial {\bf A}'_{\gamma}(66')\partial {\bf A}_{\beta}(33')}\bigg|_{{\bf A}'={\bm 0}}+\frac{\partial {\bm \Sigma}^{({\bf A}')}(22')}{\partial {\bf G}^{({\bf A}')}(44')}{\bf G}^{({\bf A}')}(45)
\frac{\partial^2 {\bf G}^{ ({\bf A}')-1}(55')}{\partial {\bf A}_{\gamma}'(66')\partial {\bf A}_{\beta}(33')}{\bf G}^{({\bf A}')}(5'4')\bigg|_{{\bf A}'={\bm 0}}\nonumber\\
&-\frac{\partial {\bm \Sigma}^{({\bf A}')}(22')}{\partial {\bf G}^{({\bf A}')}(44')}
{\bf G}^{({\bf A}')}(47)
\frac{\partial {\bf G}^{ ({\bf A}')-1}(77')}{\partial {\bf A}'_{\gamma}(66')}{\bf G}^{({\bf A}')}(7'5)
\frac{\partial {\bf G}^{ ({\bf A}')-1}(55')}{\partial {\bf A}_{\beta}(33')}{\bf G}^{({\bf A}')}(5'4')\bigg|_{{\bf A}'={\bm 0}}\nonumber
\\&-\frac{\partial {\bm \Sigma}^{({\bf A}')}(22')}{\partial {\bf G}^{({\bf A}')}(44')}{\bf G}^{({\bf A}')}(45)
\frac{\partial {\bf G}^{ ({\bf A}')-1}(55')}{\partial {\bf A}_{\beta}(33')}
{\bf G}^{({\bf A}')}(5'7)
\frac{\partial {\bf G}^{ ({\bf A}')-1}(77')}{\partial {\bf A}'_{\gamma}(66')}
{\bf G}^{({\bf A}')}(7'4')
\bigg|_{{\bf A}'={\bm 0}}\nonumber\\
&+\frac{\partial {\bm \Sigma}^{({\bf A}')}(22')}{\partial {\bf G}^{({\bf A}')}(77')\partial {\bf G}^{({\bf A}')}(44')}
\frac{\partial {\bf G}^{ ({\bf A}')}(77')}{\partial {\bf A}'_{\gamma}(66')}
{\bf G}^{({\bf A}')}(45)
\frac{\partial {\bf G}^{ ({\bf A}')-1}(55')}{\partial {\bf A}_{\beta}(33')}{\bf G}^{({\bf A}')}(5'4')\bigg|_{{\bf A}'={\bm 0}}.\label{E123}
\end{align}
\end{widetext}
The last term describes irreducible three body interactions. As mentioned above, we assume that this term can be neglected. Let us evaluate $\partial_{A_\gamma}$ of the paramagnetic term \eref{E121}. Then consider what happens when, for example, the term coming from the second line of  \eref{E123} is added to the term coming from the derivative of the first propagator in \eref{E121}. We obtain 
\begin{widetext}
\begin{align}
&-{\bm \lambda}_{\alpha}(1'1){\bf G}^{ }(17){\bm \Lambda}_{\gamma}(77') {\bf G} (7'5){\bm \Lambda}_{\beta}(55') {\bf G} (5'1')\nonumber\\
&-{\bf G} (2'1'){\bm \lambda}_{\alpha} (1'1){\bf G}(12)\frac{\partial {\bm \Sigma}(22')}{\partial {\bf G}(44')}
{\bf G}(47)
{\bm \Lambda}_{\gamma}(77'){\bf G}(7'5)
{\bm \Lambda}_{\beta}(55') {\bf G}(5'4'),
\end{align}
\end{widetext}
which can be rewritten as the first line except that the bare $\alpha$ vertex is replaced with the first order of the iterative solution of the dressed $\alpha$ vertex.

\section{Hall conductivity}\label{D}
In this section we generalize the single band formalism~\cite{10.1143/PTP.42.1284, 0305-4608-15-8-011} to the multi-band case. 
In order to obtain the Hall response, a transverse response, $\Pi^{(\eta)}_{\alpha \beta}$, must be calculated in presence of a perpendicular magnetic field, $B$. The magnetic field can be obtained from a gauge potential ${\bf B}({\bf q}',\nu') = i{\bf q}'\times {\bf A}'({\bf q}',\nu')$, where ${\bf q}',\nu'$ denote the wave vector and  the frequency of the magnetic field. That frequency is taken to zero here since the magnetic field is time independent. Therefore, the total gauge potential is ${\bf A}(t)+{\bf A}'({\bf r})$ and the Hall response is proportional to ${\bf E}\times {\bf B}=\nu\left[{\bf A}'({\bf q}' \cdot {\bf A}) - {\bf q}'({\bf A}\cdot {\bf A}') \right]$, where $\nu$ denotes the frequency of the electric field.
Once the derivative with respect to ${\bf q'}$ has been calculated one must take ${\bf q'}= {\bm 0}$ to obtain the linear response to a uniform magnetic field. 

The linear response of $\Pi^{(\eta)}_{\alpha \beta}$ to the magnetic field is calculated, as explained in \eref{eq:Hall_conductivity} of the main text, from:
%
\begin{widetext}
\begin{align}
-i\frac{\epsilon^{\eta\delta \gamma }}{2}\partial_{q'_{\delta}}\partial_{A'_{\gamma}} \Pi^{({\bf A}')}_{\alpha \beta}(i\nu_n)\bigg|_{({\bf A}',{\bf q}')={\bm 0}}=& (\frac{i\epsilon^{\eta\delta \gamma }k_BT}{2V_{\rm cell}N})\partial_{q'_{\delta}}\partial_{A'_{\gamma}}\sum_{{\bf k},\omega_m}{\rm Tr}\bigg\{ 
\nonumber \\
\left[{\bm \lambda}_{\alpha}^{({\bf A}')}({\bf q}',\nu_n)\right]_{{\bf k}'_-,\omega_m;{\bf k}'_+,\omega_m^+}
&{\bf G}^{({\bf A}')}_{{\bf k}'_+\omega_m^+;{\bf k}'_-,\omega_m^+}
\left[{\bm \Lambda}_{\beta}^{({\bf A}')}({\bf q}',-\nu_n)\right]_{{\bf k}'_-,\omega_m^+; {\bf k}'_+,\omega_m}
{\bf G}^{({\bf A}')}_{{\bf k}'_+, \omega_m;{\bf k}'_-,\omega_m}\bigg\}\bigg|_{({\bf A}',{\bf q}')={\bm 0}}. 
\end{align}
\end{widetext}
where we have defined ${\bf k}'_{\pm}\equiv {\bf k} \pm {\bf q}'/2$ and used the definition of the bare \eref{eq:bare_vertex} (\eref{eq:def_lambda}) and dressed \eref{eq:dressed_vertex} (\eref{eq:def_Lambda}) current vertices, {\it before} taking the ${\bf A}'=0$ limit. Note that the diamagnetic term is proportional to $\delta_{\alpha\beta}$ and hence does not contribute to the Hall conductivity.  By applying the derivative with respect to $A_{\gamma}$ as in the preceding appendix, then setting $A_{\gamma}=0$  and keeping only terms linear in ${\bf q}'$, we find four terms,
\begin{widetext}
\begin{align}
-i\frac{\epsilon^{\eta\delta \gamma }}{2}&\partial_{q'_{\delta}}\partial_{A'_{\gamma}} \Pi^{({\bf A}')}_{\alpha \beta}(i\nu_n)\bigg|_{({\bf A}',{\bf q}')={\bm 0}}= (\frac{i\epsilon^{\eta\delta \gamma }k_BT}{2V_{\rm cell}N})\partial_{q'_{\delta}}\sum_{{\bf k},\omega_m}{\rm Tr}\bigg\{\nonumber \\ 
&\left[{\bm \lambda}_{\gamma \alpha}({\bm 0}, \nu_n)\right]_{{\bf k}, \omega_m;{\bf k}, \omega_m^+}
{\bf G}_{{\bf k}, \omega_m^+}
\left[{\bm \Lambda}_{\beta}({\bm 0}, -\nu_n)\right]_{{\bf k}, \omega_m^+;{\bf k}, \omega_m}
{\bf G}_{{\bf k}, \omega_m}+\nonumber \\
&\left[{\bm \Lambda}_{\alpha}({\bf q}',\nu_n)\right]_{{\bf k}'_-,\omega_m;{\bf k}'_+,\omega_m^+}
{\bf G}_{{\bf k}'_+,\omega_m^+}
\left[{\bm \Lambda}_{\gamma}(-{\bf q}',0)\right]_{{\bf k}'_+,\omega_m^+;{\bf k}'_-,\omega_m^+}
{\bf G}_{{\bf k}'_-, \omega_m^+}
\left[{\bm \Lambda}_{\beta}({\bm 0},-\nu_n)\right]_{{\bf k}'_-,\omega_m^+;{\bf k}'_-,\omega_m}
{\bf G}_{{\bf k}'_-, \omega_m}+\nonumber \\
&\left[{\bm \Lambda}_{\alpha}({\bf q}', \nu_n)\right]_{{\bf k}'_-,\omega_m;{\bf k}'_+,\omega_m^+}
{\bf G}_{{\bf k}'_+, \omega_m^+}
\left[{\bm \Lambda}_{\gamma \beta}(-{\bf q}',-\nu_n)\right]_{{\bf k}'_+,\omega_m^+;{\bf k}'_-,\omega_m}
{\bf G}_{{\bf k}'_-,\omega_m}+\nonumber \\
&\left[{\bm \Lambda}_{\alpha}({\bf q}',\nu_n)\right]_{{\bf k}'_-,\omega_m;{\bf k}'_+,\omega_m^+}
{\bf G}_{{\bf k}'_+,\omega_m^+}
\left[{\bm \Lambda}_{\beta}({\bm 0},-\nu_n)\right]_{{\bf k}'_+,\omega_m^+;{\bf k}'_+,\omega_m}
{\bf G}_{{\bf k}'_+, \omega_m}
\left[{\bm \Lambda}_{\gamma}(-{\bf q}',0)\right]_{{\bf k}'_+,\omega_m;{\bf k}'_-,\omega_m}
{\bf G}_{{\bf k}'_-, \omega_m}\bigg\}\bigg|_{{\bf q}'={\bm 0}},\label{eq:AExp}
\end{align} 
\end{widetext} 
where we used $\partial_{A'_{\gamma}}{\bf G}^{({\bf A}')}=-{\bf G}^{({\bf A}')}(\partial_{A'_{\gamma}}{\bf G}^{({\bf A}')-1}){\bf G}^{({\bf A}')}$ and the definitions of the bare \eref{eq:bare_vertex} (\eref{eq:def_lambda}) and dressed \eref{eq:dressed_vertex} (\eref{eq:def_Lambda}) current vertices, this time in the ${\bf A}'={\bm 0}$ limit. The vertex that plays the role of the inverse effective-mass tensor, ${\bm \Lambda}_{\gamma \alpha}$ obeys the integral equation \eref{eq:def_dressed_effective_mass}, as shown in the previous appendix. 
%
%

The above four terms correspond to the four terms in \eref{eq:dA_paramagnetic}, written in full. To obtain the above result, we also used that to linear order in ${\bf q'}$, all Green functions are translationally invariant. Momentum ${\bf q'}$ flows in and out at vertices of the connected diagram. The vertex ${\bm \Lambda}_{\alpha}$, related to the observed current, changes both momentum and energy, while ${\bm \Lambda}_{\beta}$ vertex coupled to the uniform electric field only changes the energy (dissipative) and the ${\bm \Lambda}_{\gamma }$ vertex coupled to static magnetic field only changes the momentum (dispersive). 
%

The first term in \eref{eq:AExp} is ${\bf q}'$ independent because that momentum comes in and out of the same vertex, so there is no net ${\bf q}'$ dependence. Only the last three terms have a non-zero contribution to the Hall conductivity. Although we have written explicitly the net momentum and frequency going in the verticies in addition to the two momenta and frequency associated with the propagators attached to the vertices, there are only two independent momentum variables for the vertices at this stage. They can be chosen as the labels of the square parenthesis. Calculating the derivative with respect to $q'_{\delta}$ and taking the limit ${\bf q}'\rightarrow {\bm 0}$ yields triangular and rectangular diagrams, as we explain further below. 

The three-leg (triangular) diagrams are obtained when the derivative with respect to $q'_{\delta}$ acts on the vertices of the second and fourth terms and on the propagators of the third term in \eref{eq:AExp}.  To obtain the results reported in \eref{eq:Tri}, we have used that $\left[{\bm \lambda}_{\alpha}({\bf q}',\nu_n)\right]_{{\bf k}'_-,\omega_m;{\bf k}'_+,\omega_m^+}$ and $\left[{\bm \Lambda}_{\gamma} (-{\bf q}',0)\right]_{{\bf k}'_+,\omega_m;{\bf k}'_-,\omega_m}$ do not depend on ${\bf q}'$ linearly, as well as the following identities
%
\begin{align}
\partial_{q'_{\delta}}{\bf G}_{{\bf k}'_{\pm},\omega_m}\big|_{{\bf q}'={\bm 0}} = \pm(\frac{1}{2})\partial_{k_{\delta}}{\bf G}_{{\bf k},i\omega_m},
\label{eq:partial_q'}
\end{align}
and
\begin{align}\label{eq:derivarive_wrt_q_of_vertices}
\partial_{q'_{\delta}}\left[{\bm \Lambda}_{\beta}({\bm 0},-\nu_n)\right]_{{\bf k}'_{\pm},\omega_m^+;{\bf k}'_{\pm},\omega_m}\big|_{{\bf q}'={\bm 0}}  =\nonumber \\ \pm(\frac{1}{2})\partial_{k_{\delta}}\left[{\bm \Lambda}_{\beta}({\bm 0},-\nu_n)\right]_{{\bf k},\omega_m^+;{\bf k},\omega_m}.
\end{align}
We have also assumed that the derivative with respect to ${\bf k}$ is proportional to a derivative with respect to vector potential. This allows us to maintain the convention that greek indices for the vertices always stand for derivatives with respect to vector potential. But we must take into account that  
\begin{equation}\label{eq:dk=dA}
\partial_{k_\delta}[\,]=\frac{\hbar}{e}\partial_{A'_\delta}[\,]\, .
\end{equation}
This is true for this vertex that comes from the magnetic field because, in that case, only the current operator that comes from the Peierls substitution contributes. The DC magnetic field cannot lead to intra-atomic dipole terms. This allows us to set the contribution from 
$\left[{\bm \Lambda}_{\delta \gamma \beta}(-{\bf q}',-\nu_n)\right]_{{\bf k}'_+,\omega_m^+;{\bf k}'_-,\omega_m}$ to zero because of the antisymmetry of the Levi-Civita tensor. When all this is taken into account, there are two contributions from derivatives of the third term and they are equal to the other two contributions obtained from the derivative of the second and fourth term in \eref{eq:AExp} that give one contribution each. 
%

The contribution of the four-leg (rectangular) diagrams in \eref{eq:Rec} is obtained from derivatives with respect to ${\bf q}'$ of the propagators in the second and fourth terms of \eref{eq:AExp} and by the replacement of the derivative with respect to ${\bf q}'$ by a derivative with respect to ${\bf k}$ as in \eref{eq:partial_q'}. This gives six terms that can be combined in pairs using the antisymmetry in $\gamma,\delta$ of the Levi-Civita tensor. 
%

\section{ Hall response in the $\nu_n \rightarrow 0$ limit }\label{Sec:Appendix_DC}
\subsection{Interband triangular diagrams}
We mentioned in the main text that the interband components of the triangular diagrams do not contribute to the Hall response in the $\nu_n \rightarrow \nu_1=2\pi k_BT$ limit at low temperature. 
Here, we explain this more.  In the band basis, the interband terms can be classified in two classes in terms of the band indices of the three Green's function: (i) terms with three different band indices, and (ii) terms with two different band indices. 

The Matsubara frequency summation of the first class includes only simple poles. Therefore, the resulting expression after performing summation over fermionic Matsubara frequencies is regular in the $\nu_1\rightarrow 0$ limit, hence, one can set the bosonic frequency equal to zero directly  for these terms in \eref{eq:Tri} and see that they vanish. 

%

For the second class of diagrams the Green's function with the same band index may have (i) different bosonic Matsubara frequency  or (ii) the same bosonic Matsubara frequency. Consider case (i). The fermionic Matsubara frequency summation can be done as in \eref{eq:0}. We expand the first term on the right side in terms of $\nu_n$ and keep the terms which do not vanishes when $\nu_n \rightarrow \nu_1$. This gives two terms: one term cancels the second term on the right-hand side. The second term is $\nu_n$-independent. This term and the third term on the right hand side of \eref{eq:0} are canceled by the  analogous terms from the second line of the \eref{eq:Tri}. In the case (ii), when the Green's functions with the same band index have the same frequency dependence, then the fermionic summation includes a second order pole where again, using \eref{eq:1}, one can show that these terms also vanish in the $\nu_n \rightarrow \nu_1$ limit.

\subsection{Interband rectangular diagrams}
In the main text, we clarified which rectangular diagrams do not vanish at DC limit. Here, we consider one category of diagrams which vanishes in this limit and explain it in more details. For example, take $n\ne n'$ and consider diagrams with $\tilde{{\bm \lambda}}_{{\bf k},nn}^{\alpha}\tilde{{\bm \lambda}}_{{\bf k},n'n'}^{\gamma}$ and $\tilde{{\bm \lambda}}_{{\bf k},nn'}^{\delta}\tilde{{\bm \lambda}}_{{\bf k},n'n}^{\beta}$ or $\tilde{{\bm \lambda}}_{{\bf k},n'n}^{\delta}\tilde{{\bm \lambda}}_{{\bf k},nn'}^{\beta}$. Then, only the second and last lines of the \eref{eq:Rec} contribute in this class and their contribution is 
\begin{widetext}
\begin{align}
{\rm lim}_{\nu_n \rightarrow \nu_1}(\frac{-i\epsilon^{\eta\delta\gamma}Bk_BT}{2V_{\rm cell}N})\frac{\hbar}{e}&\sum_{{\bf k},\omega_m}\sum_{n,n'(\neq n)}\left[\tilde{{\bm \lambda}}_{{\bf k}}^{\alpha}\right]_{nn}\left[\tilde{{\bm \lambda}}_{{\bf k}}^{\gamma}\right]_{n'n'}
\left[{\bf G}_{{\bf k},\omega_m}\right]_{n'n'}\left[{\bf G}_{{\bf k},\omega_m}\right]_{n'n'}\bigg\{\nonumber \\
&\left( \left[\tilde{{\bm \lambda}}_{{\bf k}}^{\delta}\right]_{nn'}\left[\tilde{{\bm \lambda}}_{{\bf k}}^{\beta}\right]_{n'n}
\left[{\bf G}_{{\bf k},\omega_m}\right]_{nn}\left[{\bf G}_{{\bf k},\omega_m^-}\right]_{nn} -\left[\tilde{{\bm \lambda}}_{{\bf k}}^{\beta}\right]_{nn'}\left[\tilde{{\bm \lambda}}_{{\bf k}}^{\delta}\right]_{n'n}
\left[{\bf G}_{{\bf k},\omega_m}\right]_{nn}\left[{\bf G}_{{\bf k},\omega_m^+}\right]_{nn} \right) \bigg\}=0,
\end{align}
\end{widetext}
where we used \eref{eq:6} and expanded in powers of $\nu_1$. Similar calculations can be done for all the other vanishing diagrams.

\section{Matsubara frequency summations}\label{Sec:Matsubara}
Here we list some identities that have been used in this paper. In the following we consider a non-interacting system in the band basis where $E_{n,{\bf k}}$ denotes band energy. For the Green's functions below, we take the definition~\eref{eq:GreenGamma} with $\Gamma=0$. 
\begin{widetext}
\begin{align}
k_BT\sum_{\omega_m}\left[{\bf G}_{{\bf k},\omega_m^+}\right]_{n_1n_1}\left[{\bf G}_{{\bf k},\omega_m}\right]_{n_1n_1}&\left[{\bf G}_{{\bf k},\omega_m}\right]_{n_2n_2}=\frac{f(E_{n_1,{\bf k}})}{(-i\nu_n)(E_{n_1,{\bf k}}-E_{n_2,{\bf k}}-i\nu_n)}+\frac{f(E_{n_1,{\bf k}})}{(i\nu_n)(E_{n_1,{\bf k}}-E_{n_2,{\bf k}})}+\nonumber\\
&\frac{f(E_{n_2,{\bf k}})}{(E_{n_2,{\bf k}}-E_{n_1,{\bf k}}+i\nu_n)(E_{n_2,{\bf k}}-E_{n_1,{\bf k}})},\label{eq:0}\\
k_BT\sum_{\omega_m}
\left[{\bf G}_{{\bf k},\omega_m^+}\right]_{n_2n_2}
\left[{\bf G}_{{\bf k},\omega_m}\right]^2_{n_1n_1}&
=\frac{f(E_{n_2,{\bf k}})}{(E_{n_2,{\bf k}}-E_{n_1,{\bf k}}-i\nu_n)^2}+\frac{f'(E_{n_1,{\bf k}})}{(E_{n_1,{\bf k}}-E_{n_2,{\bf k}}+i\nu_n)}-\frac{f(E_{n_1,{\bf k}})}{(E_{n_1,{\bf k}}-E_{n_2,{\bf k}}+i\nu_n)^2},\label{eq:1}\\
%
k_BT\sum_{\omega_m}\left[{\bf G}_{{\bf k},\omega_m^+}\right]_{n_1n_1}\left[{\bf G}_{{\bf k},\omega_m}\right]^2_{n_1n_1}&\left[{\bf G}_{{\bf k},\omega_m}\right]_{n_2n_2}=\frac{1}{(E_{n_2,{\bf k}}-E_{n_1,{\bf k}})^2}\frac{f(E_{n_2,{\bf k}})}{(E_{n_2,{\bf k}}-E_{n_1,{\bf k}}+i\nu_n)}+
 \nonumber\\
 (\frac{1}{-i\nu_n})^2\frac{f(E_{n_1,{\bf k}})}{(E_{n_1,{\bf k}}-E_{n_2,{\bf k}}-i\nu_n)}
 &+(\frac{1}{i\nu_n})\frac{f'(E_{n_1,{\bf k}})}{(E_{n_1,{\bf k}}-E_{n_2,{\bf k}})}-(\frac{1}{i\nu_n})\frac{f(E_{n_1,{\bf k}})}{(E_{n_1,{\bf k}}-E_{n_2,{\bf k}})^2}-(\frac{1}{i\nu_n})^2\frac{f(E_{n_1,{\bf k}})}{(E_{n_1,{\bf k}}-E_{n_2,{\bf k}})},\label{eq:3}\\
k_BT\sum_{\omega_m}\left[{\bf G}_{{\bf k},\omega_m^+}\right]_{n_1n_1}\left[{\bf G}_{{\bf k},\omega_m}\right]^2_{n_1n_1}&\left[{\bf G}_{{\bf k},\omega_m^+}\right]_{n_2n_2} = \frac{1}{(E_{n_2,{\bf k}}-E_{n_1,{\bf k}})}\frac{f(E_{n_2,{\bf k}})}{(E_{n_2,{\bf k}}-E_{n_1,{\bf k}}-i\nu_n)^2}+\nonumber \\
(\frac{1}{-i\nu_n})^2\frac{f(E_{n_1,{\bf k}})}{(E_{n_1,{\bf k}}-E_{n_2,{\bf k}})}
+(\frac{1}{i\nu_n})&\frac{f'(E_{n_1,{\bf k}})}{(E_{n_1,{\bf k}}-E_{n_2,{\bf k}}+i\nu_n)}-(\frac{1}{i\nu_n})\frac{f(E_{n_1,{\bf k}})}{(E_{n_1,{\bf k}}-E_{n_2,{\bf k}}+i\nu_n)^2}-(\frac{1}{i\nu_n})^2\frac{f(E_{n_1,{\bf k}})}{(E_{n_1,{\bf k}}-E_{n_2,{\bf k}}+i\nu_n)}, \label{eq:4} \\
k_BT\sum_{\omega_m}\left[{\bf G}_{{\bf k},\omega_m^+}\right]^2_{n_1n_1}\left[{\bf G}_{{\bf k},\omega_m}\right]_{n_1n_1}&\left[{\bf G}_{{\bf k},\omega_m}\right]_{n_2n_2} = \frac{1}{(E_{n_2,{\bf k}}-E_{n_1,{\bf k}})}\frac{f(E_{n_2,{\bf k}})}{(E_{n_2,{\bf k}}-E_{n_1,{\bf k}}+i\nu_n)^2}+\nonumber \\
(\frac{1}{i\nu_n})^2\frac{f(E_{n_1,{\bf k}})}{(E_{n_1,{\bf k}}-E_{n_2,{\bf k}})}
+(\frac{1}{-i\nu_n})&\frac{f'(E_{n_1,{\bf k}})}{(E_{n_1,{\bf k}}-E_{n_2,{\bf k}}-i\nu_n)}-(\frac{1}{-i\nu_n})\frac{f(E_{n_1,{\bf k}})}{(E_{n_1,{\bf k}}-E_{n_2,{\bf k}}-i\nu_n)^2}-(\frac{1}{-i\nu_n})^2\frac{f(E_{n_1,{\bf k}})}{(E_{n_1,{\bf k}}-E_{n_2,{\bf k}}-i\nu_n)}, \label{eq:5}\\
k_BT\sum_{\omega_m}\left[{\bf G}_{{\bf k},\omega_m}\right]^2_{n_1n_1}\left[{\bf G}_{{\bf k},\omega_m}\right]_{n_2n_2}&\left[{\bf G}_{{\bf k},\omega_m^+}\right]_{n_2n_2} = \frac{1}{(i\nu_n)}\frac{f(E_{n_2,{\bf k}})}{(E_{n_2,{\bf k}}-E_{n_1,{\bf k}})^2}+
(\frac{1}{-i\nu_n})\frac{f(E_{n_2,{\bf k}})}{(E_{n_2,{\bf k}}-E_{n_1,{\bf k}}-i\nu_n)^2}
+\nonumber \\\frac{1}{(E_{n_1,{\bf k}}-E_{n_2,{\bf k}})}\frac{f'(E_{n_1,{\bf k}})}{(E_{n_1,{\bf k}}-E_{n_2,{\bf k}}+i\nu_n)}&-\frac{1}{(E_{n_1,{\bf k}}-E_{n_2,{\bf k}})^2}\frac{f(E_{n_1,{\bf k}})}{(E_{n_1,{\bf k}}-E_{n_2,{\bf k}}+i\nu_n)}-\frac{1}{(E_{n_1,{\bf k}}-E_{n_2,{\bf k}})}\frac{f(E_{n_1,{\bf k}})}{(E_{n_1,{\bf k}}-E_{n_2,{\bf k}}+i\nu_n)^2}, \label{eq:6}
\end{align}
\end{widetext}
where $f'(E_{n_1,{\bf k}}) = (\partial f(E_{n_1,{\bf k}})/\partial_{E_{n_1,{\bf k}}})$.


\begin{thebibliography}{38}%
\makeatletter
\providecommand \@ifxundefined [1]{%
 \@ifx{#1\undefined}
}%
\providecommand \@ifnum [1]{%
 \ifnum #1\expandafter \@firstoftwo
 \else \expandafter \@secondoftwo
 \fi
}%
\providecommand \@ifx [1]{%
 \ifx #1\expandafter \@firstoftwo
 \else \expandafter \@secondoftwo
 \fi
}%
\providecommand \natexlab [1]{#1}%
\providecommand \enquote  [1]{``#1''}%
\providecommand \bibnamefont  [1]{#1}%
\providecommand \bibfnamefont [1]{#1}%
\providecommand \citenamefont [1]{#1}%
\providecommand \href@noop [0]{\@secondoftwo}%
\providecommand \href [0]{\begingroup \@sanitize@url \@href}%
\providecommand \@href[1]{\@@startlink{#1}\@@href}%
\providecommand \@@href[1]{\endgroup#1\@@endlink}%
\providecommand \@sanitize@url [0]{\catcode `\\12\catcode `\$12\catcode
  `\&12\catcode `\#12\catcode `\^12\catcode `\_12\catcode `\%12\relax}%
\providecommand \@@startlink[1]{}%
\providecommand \@@endlink[0]{}%
\providecommand \url  [0]{\begingroup\@sanitize@url \@url }%
\providecommand \@url [1]{\endgroup\@href {#1}{\urlprefix }}%
\providecommand \urlprefix  [0]{URL }%
\providecommand \Eprint [0]{\href }%
\providecommand \doibase [0]{http://dx.doi.org/}%
\providecommand \selectlanguage [0]{\@gobble}%
\providecommand \bibinfo  [0]{\@secondoftwo}%
\providecommand \bibfield  [0]{\@secondoftwo}%
\providecommand \translation [1]{[#1]}%
\providecommand \BibitemOpen [0]{}%
\providecommand \bibitemStop [0]{}%
\providecommand \bibitemNoStop [0]{.\EOS\space}%
\providecommand \EOS [0]{\spacefactor3000\relax}%
\providecommand \BibitemShut  [1]{\csname bibitem#1\endcsname}%
\let\auto@bib@innerbib\@empty
\bibitem [{\citenamefont {Nair}\ \emph {et~al.}(2012)\citenamefont {Nair},
  \citenamefont {Wirth}, \citenamefont {Friedemann}, \citenamefont {Steglich},
  \citenamefont {Si},\ and\ \citenamefont
  {Schofield}}]{Nair_Wirth_Friedemann_Steglich_Si_Schofield_2012}%
  \BibitemOpen
  \bibfield  {author} {\bibinfo {author} {\bibfnamefont {S.}~\bibnamefont
  {Nair}}, \bibinfo {author} {\bibfnamefont {S.}~\bibnamefont {Wirth}},
  \bibinfo {author} {\bibfnamefont {S.}~\bibnamefont {Friedemann}}, \bibinfo
  {author} {\bibfnamefont {F.}~\bibnamefont {Steglich}}, \bibinfo {author}
  {\bibfnamefont {Q.}~\bibnamefont {Si}}, \ and\ \bibinfo {author}
  {\bibfnamefont {A.~J.}\ \bibnamefont {Schofield}},\ }\href {\doibase
  10.1080/00018732.2012.730223} {\bibfield  {journal} {\bibinfo  {journal}
  {Advances in Physics}\ }\textbf {\bibinfo {volume} {61}},\ \bibinfo {pages}
  {583–664} (\bibinfo {year} {2012})}\BibitemShut {NoStop}%
\bibitem [{\citenamefont {Hall}(1879)}]{Hall:1879}%
  \BibitemOpen
  \bibfield  {author} {\bibinfo {author} {\bibfnamefont {E.~H.}\ \bibnamefont
  {Hall}},\ }\href {http://www.jstor.org/stable/2369245} {\bibfield  {journal}
  {\bibinfo  {journal} {American Journal of Mathematics}\ }\textbf {\bibinfo
  {volume} {2}},\ \bibinfo {pages} {287} (\bibinfo {year} {1879})}\BibitemShut
  {NoStop}%
\bibitem [{\citenamefont {Shastry}\ \emph {et~al.}(1993)\citenamefont
  {Shastry}, \citenamefont {Shraiman},\ and\ \citenamefont
  {Singh}}]{PhysRevLett.70.2004}%
  \BibitemOpen
  \bibfield  {author} {\bibinfo {author} {\bibfnamefont {B.~S.}\ \bibnamefont
  {Shastry}}, \bibinfo {author} {\bibfnamefont {B.~I.}\ \bibnamefont
  {Shraiman}}, \ and\ \bibinfo {author} {\bibfnamefont {R.~R.~P.}\ \bibnamefont
  {Singh}},\ }\href {\doibase 10.1103/PhysRevLett.70.2004} {\bibfield
  {journal} {\bibinfo  {journal} {Phys. Rev. Lett.}\ }\textbf {\bibinfo
  {volume} {70}},\ \bibinfo {pages} {2004} (\bibinfo {year}
  {1993})}\BibitemShut {NoStop}%
\bibitem [{\citenamefont {Basov}\ \emph {et~al.}(2011)\citenamefont {Basov},
  \citenamefont {Averitt}, \citenamefont {van~der Marel}, \citenamefont
  {Dressel},\ and\ \citenamefont {Haule}}]{RevModPhys.83.471}%
  \BibitemOpen
  \bibfield  {author} {\bibinfo {author} {\bibfnamefont {D.}~\bibnamefont
  {Basov}}, \bibinfo {author} {\bibfnamefont {R.}~\bibnamefont {Averitt}},
  \bibinfo {author} {\bibfnamefont {D.}~\bibnamefont {van~der Marel}}, \bibinfo
  {author} {\bibfnamefont {M.}~\bibnamefont {Dressel}}, \ and\ \bibinfo
  {author} {\bibfnamefont {K.}~\bibnamefont {Haule}},\ }\href {\doibase
  10.1103/RevModPhys.83.471} {\bibfield  {journal} {\bibinfo  {journal} {Rev.
  Mod. Phys.}\ }\textbf {\bibinfo {volume} {83}},\ \bibinfo {pages} {471}
  (\bibinfo {year} {2011})}\BibitemShut {NoStop}%
\bibitem [{\citenamefont {Badoux}\ \emph {et~al.}(2016)\citenamefont {Badoux},
  \citenamefont {Tabis}, \citenamefont {Lalibert{\'e}}, \citenamefont
  {Grissonnanche}, \citenamefont {Vignolle}, \citenamefont {Vignolles},
  \citenamefont {B{\'e}ard}, \citenamefont {Bonn}, \citenamefont {Hardy},
  \citenamefont {Liang} \emph {et~al.}}]{Nature.10.1038/nature16983}%
  \BibitemOpen
  \bibfield  {author} {\bibinfo {author} {\bibfnamefont {S.}~\bibnamefont
  {Badoux}}, \bibinfo {author} {\bibfnamefont {W.}~\bibnamefont {Tabis}},
  \bibinfo {author} {\bibfnamefont {F.}~\bibnamefont {Lalibert{\'e}}}, \bibinfo
  {author} {\bibfnamefont {G.}~\bibnamefont {Grissonnanche}}, \bibinfo {author}
  {\bibfnamefont {B.}~\bibnamefont {Vignolle}}, \bibinfo {author}
  {\bibfnamefont {D.}~\bibnamefont {Vignolles}}, \bibinfo {author}
  {\bibfnamefont {J.}~\bibnamefont {B{\'e}ard}}, \bibinfo {author}
  {\bibfnamefont {D.}~\bibnamefont {Bonn}}, \bibinfo {author} {\bibfnamefont
  {W.}~\bibnamefont {Hardy}}, \bibinfo {author} {\bibfnamefont
  {R.}~\bibnamefont {Liang}},  \emph {et~al.},\ }\href {\doibase
  10.1038/nature16983} {\bibfield  {journal} {\bibinfo  {journal} {Nature}\ }
  (\bibinfo {year} {2016}),\ 10.1038/nature16983}\BibitemShut {NoStop}%
\bibitem [{\citenamefont {Collignon}\ \emph {et~al.}(2017)\citenamefont
  {Collignon}, \citenamefont {Badoux}, \citenamefont {Afshar}, \citenamefont
  {Michon}, \citenamefont {Lalibert\'e}, \citenamefont {Cyr-Choini\`ere},
  \citenamefont {Zhou}, \citenamefont {Licciardello}, \citenamefont {Wiedmann},
  \citenamefont {Doiron-Leyraud},\ and\ \citenamefont
  {Taillefer}}]{PhysRevB.95.224517}%
  \BibitemOpen
  \bibfield  {author} {\bibinfo {author} {\bibfnamefont {C.}~\bibnamefont
  {Collignon}}, \bibinfo {author} {\bibfnamefont {S.}~\bibnamefont {Badoux}},
  \bibinfo {author} {\bibfnamefont {S.~A.~A.}\ \bibnamefont {Afshar}}, \bibinfo
  {author} {\bibfnamefont {B.}~\bibnamefont {Michon}}, \bibinfo {author}
  {\bibfnamefont {F.}~\bibnamefont {Lalibert\'e}}, \bibinfo {author}
  {\bibfnamefont {O.}~\bibnamefont {Cyr-Choini\`ere}}, \bibinfo {author}
  {\bibfnamefont {J.-S.}\ \bibnamefont {Zhou}}, \bibinfo {author}
  {\bibfnamefont {S.}~\bibnamefont {Licciardello}}, \bibinfo {author}
  {\bibfnamefont {S.}~\bibnamefont {Wiedmann}}, \bibinfo {author}
  {\bibfnamefont {N.}~\bibnamefont {Doiron-Leyraud}}, \ and\ \bibinfo {author}
  {\bibfnamefont {L.}~\bibnamefont {Taillefer}},\ }\href {\doibase
  10.1103/PhysRevB.95.224517} {\bibfield  {journal} {\bibinfo  {journal} {Phys.
  Rev. B}\ }\textbf {\bibinfo {volume} {95}},\ \bibinfo {pages} {224517}
  (\bibinfo {year} {2017})}\BibitemShut {NoStop}%
\bibitem [{\citenamefont {Fanfarillo}\ \emph {et~al.}(2012)\citenamefont
  {Fanfarillo}, \citenamefont {Cappelluti}, \citenamefont {Castellani},\ and\
  \citenamefont {Benfatto}}]{PhysRevLett.109.096402}%
  \BibitemOpen
  \bibfield  {author} {\bibinfo {author} {\bibfnamefont {L.}~\bibnamefont
  {Fanfarillo}}, \bibinfo {author} {\bibfnamefont {E.}~\bibnamefont
  {Cappelluti}}, \bibinfo {author} {\bibfnamefont {C.}~\bibnamefont
  {Castellani}}, \ and\ \bibinfo {author} {\bibfnamefont {L.}~\bibnamefont
  {Benfatto}},\ }\href {\doibase 10.1103/PhysRevLett.109.096402} {\bibfield
  {journal} {\bibinfo  {journal} {Phys. Rev. Lett.}\ }\textbf {\bibinfo
  {volume} {109}},\ \bibinfo {pages} {096402} (\bibinfo {year}
  {2012})}\BibitemShut {NoStop}%
\bibitem [{\citenamefont {Kharitonov}\ \emph {et~al.}(2016)\citenamefont
  {Kharitonov}, \citenamefont {Juergens},\ and\ \citenamefont
  {Trauzettel}}]{Hall_topology_Luttinger}%
  \BibitemOpen
  \bibfield  {author} {\bibinfo {author} {\bibfnamefont {M.}~\bibnamefont
  {Kharitonov}}, \bibinfo {author} {\bibfnamefont {S.}~\bibnamefont
  {Juergens}}, \ and\ \bibinfo {author} {\bibfnamefont {B.}~\bibnamefont
  {Trauzettel}},\ }\href {\doibase 10.1103/PhysRevB.94.035146} {\bibfield
  {journal} {\bibinfo  {journal} {Phys. Rev. B}\ }\textbf {\bibinfo {volume}
  {94}},\ \bibinfo {pages} {035146} (\bibinfo {year} {2016})}\BibitemShut
  {NoStop}%
\bibitem [{\citenamefont {Zhang}\ \emph {et~al.}(2017)\citenamefont {Zhang},
  \citenamefont {Narayan}, \citenamefont {Lu}, \citenamefont {Zhang},
  \citenamefont {Zhang}, \citenamefont {Ni}, \citenamefont {Yuan},
  \citenamefont {Liu}, \citenamefont {Park}, \citenamefont {Zhang},\ and\
  \citenamefont {et~al.}}]{Zhang_Narayan:2017}%
  \BibitemOpen
  \bibfield  {author} {\bibinfo {author} {\bibfnamefont {C.}~\bibnamefont
  {Zhang}}, \bibinfo {author} {\bibfnamefont {A.}~\bibnamefont {Narayan}},
  \bibinfo {author} {\bibfnamefont {S.}~\bibnamefont {Lu}}, \bibinfo {author}
  {\bibfnamefont {J.}~\bibnamefont {Zhang}}, \bibinfo {author} {\bibfnamefont
  {H.}~\bibnamefont {Zhang}}, \bibinfo {author} {\bibfnamefont
  {Z.}~\bibnamefont {Ni}}, \bibinfo {author} {\bibfnamefont {X.}~\bibnamefont
  {Yuan}}, \bibinfo {author} {\bibfnamefont {Y.}~\bibnamefont {Liu}}, \bibinfo
  {author} {\bibfnamefont {J.-H.}\ \bibnamefont {Park}}, \bibinfo {author}
  {\bibfnamefont {E.}~\bibnamefont {Zhang}}, \ and\ \bibinfo {author}
  {\bibnamefont {et~al.}},\ }\href {\doibase 10.1038/s41467-017-01438-y}
  {\bibfield  {journal} {\bibinfo  {journal} {Nature Communications}\ }\textbf
  {\bibinfo {volume} {8}},\ \bibinfo {pages} {1272} (\bibinfo {year}
  {2017})}\BibitemShut {NoStop}%
\bibitem [{\citenamefont {Voruganti}\ \emph {et~al.}(1992)\citenamefont
  {Voruganti}, \citenamefont {Golubentsev},\ and\ \citenamefont
  {John}}]{PhysRevB.45.13945}%
  \BibitemOpen
  \bibfield  {author} {\bibinfo {author} {\bibfnamefont {P.}~\bibnamefont
  {Voruganti}}, \bibinfo {author} {\bibfnamefont {A.}~\bibnamefont
  {Golubentsev}}, \ and\ \bibinfo {author} {\bibfnamefont {S.}~\bibnamefont
  {John}},\ }\href {\doibase 10.1103/PhysRevB.45.13945} {\bibfield  {journal}
  {\bibinfo  {journal} {Phys. Rev. B}\ }\textbf {\bibinfo {volume} {45}},\
  \bibinfo {pages} {13945} (\bibinfo {year} {1992})}\BibitemShut {NoStop}%
\bibitem [{\citenamefont {Kohno}\ and\ \citenamefont
  {Yamada}(1988)}]{Prog.Theor.Phys.80.623}%
  \BibitemOpen
  \bibfield  {author} {\bibinfo {author} {\bibfnamefont {H.}~\bibnamefont
  {Kohno}}\ and\ \bibinfo {author} {\bibfnamefont {K.}~\bibnamefont {Yamada}},\
  }\href {\doibase 10.1143/PTP.80.623} {\bibfield  {journal} {\bibinfo
  {journal} {Prog. Theor. Phys.}\ }\textbf {\bibinfo {volume} {80}},\ \bibinfo
  {pages} {623} (\bibinfo {year} {1988})}\BibitemShut {NoStop}%
\bibitem [{\citenamefont {Anisimov}\ \emph {et~al.}(1997)\citenamefont
  {Anisimov}, \citenamefont {Poteryaev}, \citenamefont {Korotin}, \citenamefont
  {Anokhin},\ and\ \citenamefont
  {Kotliar}}]{Anisimov_Poteryaev_Korotin_Anokhin_Kotliar_1997}%
  \BibitemOpen
  \bibfield  {author} {\bibinfo {author} {\bibfnamefont {V.~I.}\ \bibnamefont
  {Anisimov}}, \bibinfo {author} {\bibfnamefont {A.~I.}\ \bibnamefont
  {Poteryaev}}, \bibinfo {author} {\bibfnamefont {M.~A.}\ \bibnamefont
  {Korotin}}, \bibinfo {author} {\bibfnamefont {A.~O.}\ \bibnamefont
  {Anokhin}}, \ and\ \bibinfo {author} {\bibfnamefont {G.}~\bibnamefont
  {Kotliar}},\ }\href {\doibase 10.1088/0953-8984/9/35/010} {\bibfield
  {journal} {\bibinfo  {journal} {Journal of Physics: Condensed Matter}\
  }\textbf {\bibinfo {volume} {9}},\ \bibinfo {pages} {7359} (\bibinfo {year}
  {1997})}\BibitemShut {NoStop}%
\bibitem [{\citenamefont {Kotliar}\ \emph {et~al.}(2006)\citenamefont
  {Kotliar}, \citenamefont {Savrasov}, \citenamefont {Haule}, \citenamefont
  {Oudovenko}, \citenamefont {Parcollet},\ and\ \citenamefont
  {Marianetti}}]{KotliarRMP:2006}%
  \BibitemOpen
  \bibfield  {author} {\bibinfo {author} {\bibfnamefont {G.}~\bibnamefont
  {Kotliar}}, \bibinfo {author} {\bibfnamefont {S.~Y.}\ \bibnamefont
  {Savrasov}}, \bibinfo {author} {\bibfnamefont {K.}~\bibnamefont {Haule}},
  \bibinfo {author} {\bibfnamefont {V.~S.}\ \bibnamefont {Oudovenko}}, \bibinfo
  {author} {\bibfnamefont {O.}~\bibnamefont {Parcollet}}, \ and\ \bibinfo
  {author} {\bibfnamefont {C.~A.}\ \bibnamefont {Marianetti}},\ }\href
  {\doibase 10.1103/RevModPhys.78.865} {\bibfield  {journal} {\bibinfo
  {journal} {Reviews of Modern Physics}\ }\textbf {\bibinfo {volume} {78}},\
  \bibinfo {eid} {865} (\bibinfo {year} {2006})}\BibitemShut {NoStop}%
\bibitem [{\citenamefont {Pavarini}\ \emph {et~al.}(2011)\citenamefont
  {Pavarini}, \citenamefont {Koch}, \citenamefont {Lichtenstein},\ and\
  \citenamefont {Vollhardt}}]{Pavarini:17645}%
  \BibitemOpen
  \bibfield  {author} {\bibinfo {author} {\bibfnamefont {E.}~\bibnamefont
  {Pavarini}}, \bibinfo {author} {\bibfnamefont {E.}~\bibnamefont {Koch}},
  \bibinfo {author} {\bibfnamefont {A.}~\bibnamefont {Lichtenstein}}, \ and\
  \bibinfo {author} {\bibfnamefont {D.~E.}\ \bibnamefont {Vollhardt}},\ }\href
  {http://juser.fz-juelich.de/record/17645} {\emph {\bibinfo {title} {{T}he
  {LDA}+{DMFT} approach to strongly correlated materials}}},\ \bibinfo {series}
  {Schriften des Forschungszentrums Jülich : Modeling and Simulation},
  Vol.~\bibinfo {volume} {1}\ (\bibinfo {year} {2011})\ \bibinfo {note} {record
  converted from VDB: 12.11.2012}\BibitemShut {NoStop}%
\bibitem [{\citenamefont {Aichhorn}\ \emph {et~al.}(2015)\citenamefont
  {Aichhorn}, \citenamefont {Pourovskii}, \citenamefont {Seth}, \citenamefont
  {Vildosola}, \citenamefont {Zingl}, \citenamefont {Peil}, \citenamefont
  {Deng}, \citenamefont {Mravlje}, \citenamefont {Kraberger}, \citenamefont
  {Martins},\ and\ \citenamefont {et~al.}}]{Aichhorn_Pourovskii:2015}%
  \BibitemOpen
  \bibfield  {author} {\bibinfo {author} {\bibfnamefont {M.}~\bibnamefont
  {Aichhorn}}, \bibinfo {author} {\bibfnamefont {L.}~\bibnamefont
  {Pourovskii}}, \bibinfo {author} {\bibfnamefont {P.}~\bibnamefont {Seth}},
  \bibinfo {author} {\bibfnamefont {V.}~\bibnamefont {Vildosola}}, \bibinfo
  {author} {\bibfnamefont {M.}~\bibnamefont {Zingl}}, \bibinfo {author}
  {\bibfnamefont {O.~E.}\ \bibnamefont {Peil}}, \bibinfo {author}
  {\bibfnamefont {X.}~\bibnamefont {Deng}}, \bibinfo {author} {\bibfnamefont
  {J.}~\bibnamefont {Mravlje}}, \bibinfo {author} {\bibfnamefont {G.~J.}\
  \bibnamefont {Kraberger}}, \bibinfo {author} {\bibfnamefont {C.}~\bibnamefont
  {Martins}}, \ and\ \bibinfo {author} {\bibnamefont {et~al.}},\ }\href
  {http://arxiv.org/abs/1511.01302} {\bibfield  {journal} {\bibinfo  {journal}
  {arXiv:1511.01302 [cond-mat]}\ } (\bibinfo {year} {2015})},\ \bibinfo {note}
  {arXiv: 1511.01302}\BibitemShut {NoStop}%
\bibitem [{\citenamefont {Galler}\ \emph {et~al.}(2017)\citenamefont {Galler},
  \citenamefont {Kaufmann}, \citenamefont {Gunacker}, \citenamefont
  {Thunstrom}, \citenamefont {Tomczak},\ and\ \citenamefont
  {Held}}]{Galler_Kaufmann_Gunacker_Thunstrom_Tomczak_Held_2017}%
  \BibitemOpen
  \bibfield  {author} {\bibinfo {author} {\bibfnamefont {A.}~\bibnamefont
  {Galler}}, \bibinfo {author} {\bibfnamefont {J.}~\bibnamefont {Kaufmann}},
  \bibinfo {author} {\bibfnamefont {P.}~\bibnamefont {Gunacker}}, \bibinfo
  {author} {\bibfnamefont {P.}~\bibnamefont {Thunstrom}}, \bibinfo {author}
  {\bibfnamefont {J.~M.}\ \bibnamefont {Tomczak}}, \ and\ \bibinfo {author}
  {\bibfnamefont {K.}~\bibnamefont {Held}},\ }\href
  {http://arxiv.org/abs/1709.02663} {\bibfield  {journal} {\bibinfo  {journal}
  {arXiv:1709.02663 [cond-mat]}\ } (\bibinfo {year} {2017})},\ \bibinfo {note}
  {arXiv: 1709.02663}\BibitemShut {NoStop}%
\bibitem [{\citenamefont {Marzari}\ \emph {et~al.}(2012)\citenamefont
  {Marzari}, \citenamefont {Mostofi}, \citenamefont {Yates}, \citenamefont
  {Souza},\ and\ \citenamefont {Vanderbilt}}]{RevModPhys.84.1419}%
  \BibitemOpen
  \bibfield  {author} {\bibinfo {author} {\bibfnamefont {N.}~\bibnamefont
  {Marzari}}, \bibinfo {author} {\bibfnamefont {A.~A.}\ \bibnamefont
  {Mostofi}}, \bibinfo {author} {\bibfnamefont {J.~R.}\ \bibnamefont {Yates}},
  \bibinfo {author} {\bibfnamefont {I.}~\bibnamefont {Souza}}, \ and\ \bibinfo
  {author} {\bibfnamefont {D.}~\bibnamefont {Vanderbilt}},\ }\href {\doibase
  10.1103/RevModPhys.84.1419} {\bibfield  {journal} {\bibinfo  {journal} {Rev.
  Mod. Phys.}\ }\textbf {\bibinfo {volume} {84}},\ \bibinfo {pages} {1419}
  (\bibinfo {year} {2012})}\BibitemShut {NoStop}%
\bibitem [{\citenamefont {Paul}\ and\ \citenamefont
  {Kotliar}(2003)}]{Paul_Kotliar:2003}%
  \BibitemOpen
  \bibfield  {author} {\bibinfo {author} {\bibfnamefont {I.}~\bibnamefont
  {Paul}}\ and\ \bibinfo {author} {\bibfnamefont {G.}~\bibnamefont {Kotliar}},\
  }\href {\doibase 10.1103/PhysRevB.67.115131} {\bibfield  {journal} {\bibinfo
  {journal} {Phys. Rev. B}\ }\textbf {\bibinfo {volume} {67}},\ \bibinfo
  {pages} {115131} (\bibinfo {year} {2003})}\BibitemShut {NoStop}%
\bibitem [{\citenamefont {Ogata}\ and\ \citenamefont
  {Fukuyama}(2015)}]{doi:10.7566/JPSJ.84.124708}%
  \BibitemOpen
  \bibfield  {author} {\bibinfo {author} {\bibfnamefont {M.}~\bibnamefont
  {Ogata}}\ and\ \bibinfo {author} {\bibfnamefont {H.}~\bibnamefont
  {Fukuyama}},\ }\href {\doibase 10.7566/JPSJ.84.124708} {\bibfield  {journal}
  {\bibinfo  {journal} {Journal of the Physical Society of Japan}\ }\textbf
  {\bibinfo {volume} {84}},\ \bibinfo {pages} {124708} (\bibinfo {year}
  {2015})},\ \Eprint
  {http://arxiv.org/abs/http://dx.doi.org/10.7566/JPSJ.84.124708}
  {http://dx.doi.org/10.7566/JPSJ.84.124708} \BibitemShut {NoStop}%
\bibitem [{\citenamefont {Tomczak}\ and\ \citenamefont
  {Biermann}(2009)}]{PhysRevB.80.085117}%
  \BibitemOpen
  \bibfield  {author} {\bibinfo {author} {\bibfnamefont {J.~M.}\ \bibnamefont
  {Tomczak}}\ and\ \bibinfo {author} {\bibfnamefont {S.}~\bibnamefont
  {Biermann}},\ }\href {\doibase 10.1103/PhysRevB.80.085117} {\bibfield
  {journal} {\bibinfo  {journal} {Phys. Rev. B}\ }\textbf {\bibinfo {volume}
  {80}},\ \bibinfo {pages} {085117} (\bibinfo {year} {2009})}\BibitemShut
  {NoStop}%
\bibitem [{\citenamefont {Mahan}(2000)}]{Mahan}%
  \BibitemOpen
  \bibfield  {author} {\bibinfo {author} {\bibfnamefont {G.~D.}\ \bibnamefont
  {Mahan}},\ }\enquote {\bibinfo {title} {Many-particle physics},}\ \ (\bibinfo
   {publisher} {Kluwer Academic/Plenum Publishers},\ \bibinfo {address} {New
  York},\ \bibinfo {year} {2000})\BibitemShut {NoStop}%
\bibitem [{\citenamefont {Nourafkan}\ \emph {et~al.}(2014)\citenamefont
  {Nourafkan}, \citenamefont {Kotliar},\ and\ \citenamefont
  {Tremblay}}]{PhysRevB.90.125132}%
  \BibitemOpen
  \bibfield  {author} {\bibinfo {author} {\bibfnamefont {R.}~\bibnamefont
  {Nourafkan}}, \bibinfo {author} {\bibfnamefont {G.}~\bibnamefont {Kotliar}},
  \ and\ \bibinfo {author} {\bibfnamefont {A.-M.~S.}\ \bibnamefont
  {Tremblay}},\ }\href {\doibase 10.1103/PhysRevB.90.125132} {\bibfield
  {journal} {\bibinfo  {journal} {Phys. Rev. B}\ }\textbf {\bibinfo {volume}
  {90}},\ \bibinfo {pages} {125132} (\bibinfo {year} {2014})}\BibitemShut
  {NoStop}%
\bibitem [{\citenamefont {Arakawa}(2016)}]{PhysRevB.94.045107}%
  \BibitemOpen
  \bibfield  {author} {\bibinfo {author} {\bibfnamefont {N.}~\bibnamefont
  {Arakawa}},\ }\href {\doibase 10.1103/PhysRevB.94.045107} {\bibfield
  {journal} {\bibinfo  {journal} {Phys. Rev. B}\ }\textbf {\bibinfo {volume}
  {94}},\ \bibinfo {pages} {045107} (\bibinfo {year} {2016})}\BibitemShut
  {NoStop}%
\bibitem [{\citenamefont {Nourafkan}\ and\ \citenamefont
  {Kotliar}(2013)}]{PhysRevB.88.155121}%
  \BibitemOpen
  \bibfield  {author} {\bibinfo {author} {\bibfnamefont {R.}~\bibnamefont
  {Nourafkan}}\ and\ \bibinfo {author} {\bibfnamefont {G.}~\bibnamefont
  {Kotliar}},\ }\href {\doibase 10.1103/PhysRevB.88.155121} {\bibfield
  {journal} {\bibinfo  {journal} {Phys. Rev. B}\ }\textbf {\bibinfo {volume}
  {88}},\ \bibinfo {pages} {155121} (\bibinfo {year} {2013})}\BibitemShut
  {NoStop}%
\bibitem [{\citenamefont {Nagaosa}\ \emph {et~al.}(2010)\citenamefont
  {Nagaosa}, \citenamefont {Sinova}, \citenamefont {Onoda}, \citenamefont
  {MacDonald},\ and\ \citenamefont {Ong}}]{RevModPhys.82.1539}%
  \BibitemOpen
  \bibfield  {author} {\bibinfo {author} {\bibfnamefont {N.}~\bibnamefont
  {Nagaosa}}, \bibinfo {author} {\bibfnamefont {J.}~\bibnamefont {Sinova}},
  \bibinfo {author} {\bibfnamefont {S.}~\bibnamefont {Onoda}}, \bibinfo
  {author} {\bibfnamefont {A.~H.}\ \bibnamefont {MacDonald}}, \ and\ \bibinfo
  {author} {\bibfnamefont {N.~P.}\ \bibnamefont {Ong}},\ }\href {\doibase
  10.1103/RevModPhys.82.1539} {\bibfield  {journal} {\bibinfo  {journal} {Rev.
  Mod. Phys.}\ }\textbf {\bibinfo {volume} {82}},\ \bibinfo {pages} {1539}
  (\bibinfo {year} {2010})}\BibitemShut {NoStop}%
\bibitem [{\citenamefont {Mitscherling}\ and\ \citenamefont
  {Metzner}(2018)}]{1807.10252}%
  \BibitemOpen
  \bibfield  {author} {\bibinfo {author} {\bibfnamefont {J.}~\bibnamefont
  {Mitscherling}}\ and\ \bibinfo {author} {\bibfnamefont {W.}~\bibnamefont
  {Metzner}},\ }\href@noop {} {\enquote {\bibinfo {title} {Longitudinal
  conductivity and hall coefficient in two-dimensional metals with spiral
  magnetic order},}\ } (\bibinfo {year} {2018}),\ \Eprint
  {http://arxiv.org/abs/arXiv:1807.10252} {arXiv:1807.10252} \BibitemShut
  {NoStop}%
\bibitem [{\citenamefont {Pugh}(1930)}]{PhysRev.36.1503}%
  \BibitemOpen
  \bibfield  {author} {\bibinfo {author} {\bibfnamefont {E.~M.}\ \bibnamefont
  {Pugh}},\ }\href {\doibase 10.1103/PhysRev.36.1503} {\bibfield  {journal}
  {\bibinfo  {journal} {Phys. Rev.}\ }\textbf {\bibinfo {volume} {36}},\
  \bibinfo {pages} {1503} (\bibinfo {year} {1930})}\BibitemShut {NoStop}%
\bibitem [{\citenamefont {Storey}(2016)}]{storey_hall_2016}%
  \BibitemOpen
  \bibfield  {author} {\bibinfo {author} {\bibfnamefont {J.~G.}\ \bibnamefont
  {Storey}},\ }\href {\doibase 10.1209/0295-5075/113/27003} {\bibfield
  {journal} {\bibinfo  {journal} {EPL (Europhysics Letters)}\ }\textbf
  {\bibinfo {volume} {113}},\ \bibinfo {pages} {27003} (\bibinfo {year}
  {2016})}\BibitemShut {NoStop}%
\bibitem [{\citenamefont {Eberlein}\ \emph {et~al.}(2016)\citenamefont
  {Eberlein}, \citenamefont {Metzner}, \citenamefont {Sachdev},\ and\
  \citenamefont {Yamase}}]{eberlein_fermi_2016-1}%
  \BibitemOpen
  \bibfield  {author} {\bibinfo {author} {\bibfnamefont {A.}~\bibnamefont
  {Eberlein}}, \bibinfo {author} {\bibfnamefont {W.}~\bibnamefont {Metzner}},
  \bibinfo {author} {\bibfnamefont {S.}~\bibnamefont {Sachdev}}, \ and\
  \bibinfo {author} {\bibfnamefont {H.}~\bibnamefont {Yamase}},\ }\href
  {\doibase 10.1103/PhysRevLett.117.187001} {\bibfield  {journal} {\bibinfo
  {journal} {Physical Review Letters}\ }\textbf {\bibinfo {volume} {117}},\
  \bibinfo {pages} {187001} (\bibinfo {year} {2016})}\BibitemShut {NoStop}%
\bibitem [{\citenamefont {Chatterjee}\ and\ \citenamefont
  {Sachdev}(2016)}]{Chatterjee_Sachdev_2016}%
  \BibitemOpen
  \bibfield  {author} {\bibinfo {author} {\bibfnamefont {S.}~\bibnamefont
  {Chatterjee}}\ and\ \bibinfo {author} {\bibfnamefont {S.}~\bibnamefont
  {Sachdev}},\ }\href {\doibase 10.1103/PhysRevB.94.205117} {\bibfield
  {journal} {\bibinfo  {journal} {Physical Review B}\ }\textbf {\bibinfo
  {volume} {94}},\ \bibinfo {pages} {205117} (\bibinfo {year}
  {2016})}\BibitemShut {NoStop}%
\bibitem [{\citenamefont {Verret}\ \emph {et~al.}(2017)\citenamefont {Verret},
  \citenamefont {Simard}, \citenamefont {Charlebois}, \citenamefont
  {S\'en\'echal},\ and\ \citenamefont
  {Tremblay}}]{verret_phenomenological_2017}%
  \BibitemOpen
  \bibfield  {author} {\bibinfo {author} {\bibfnamefont {S.}~\bibnamefont
  {Verret}}, \bibinfo {author} {\bibfnamefont {O.}~\bibnamefont {Simard}},
  \bibinfo {author} {\bibfnamefont {M.}~\bibnamefont {Charlebois}}, \bibinfo
  {author} {\bibfnamefont {D.}~\bibnamefont {S\'en\'echal}}, \ and\ \bibinfo
  {author} {\bibfnamefont {A.-M.~S.}\ \bibnamefont {Tremblay}},\ }\href
  {\doibase 10.1103/PhysRevB.96.125139} {\bibfield  {journal} {\bibinfo
  {journal} {Phys. Rev. B}\ }\textbf {\bibinfo {volume} {96}},\ \bibinfo
  {pages} {125139} (\bibinfo {year} {2017})}\BibitemShut {NoStop}%
\bibitem [{\citenamefont {Morice}\ \emph {et~al.}(2017)\citenamefont {Morice},
  \citenamefont {Montiel},\ and\ \citenamefont {P\'epin}}]{Morice-Pepin:2017}%
  \BibitemOpen
  \bibfield  {author} {\bibinfo {author} {\bibfnamefont {C.}~\bibnamefont
  {Morice}}, \bibinfo {author} {\bibfnamefont {X.}~\bibnamefont {Montiel}}, \
  and\ \bibinfo {author} {\bibfnamefont {C.}~\bibnamefont {P\'epin}},\ }\href
  {\doibase 10.1103/PhysRevB.96.134511} {\bibfield  {journal} {\bibinfo
  {journal} {Phys. Rev. B}\ }\textbf {\bibinfo {volume} {96}},\ \bibinfo
  {pages} {134511} (\bibinfo {year} {2017})}\BibitemShut {NoStop}%
\bibitem [{\citenamefont {Charlebois}\ \emph {et~al.}(2017)\citenamefont
  {Charlebois}, \citenamefont {Verret}, \citenamefont {Foley}, \citenamefont
  {Simard}, \citenamefont {Sénéchal},\ and\ \citenamefont
  {Tremblay}}]{Charlebois_Verret:2017}%
  \BibitemOpen
  \bibfield  {author} {\bibinfo {author} {\bibfnamefont {M.}~\bibnamefont
  {Charlebois}}, \bibinfo {author} {\bibfnamefont {S.}~\bibnamefont {Verret}},
  \bibinfo {author} {\bibfnamefont {A.}~\bibnamefont {Foley}}, \bibinfo
  {author} {\bibfnamefont {O.}~\bibnamefont {Simard}}, \bibinfo {author}
  {\bibfnamefont {D.}~\bibnamefont {Sénéchal}}, \ and\ \bibinfo {author}
  {\bibfnamefont {A.-M.~S.}\ \bibnamefont {Tremblay}},\ }\href {\doibase
  10.1103/PhysRevB.96.205132} {\bibfield  {journal} {\bibinfo  {journal}
  {Physical Review B}\ }\textbf {\bibinfo {volume} {96}} (\bibinfo {year}
  {2017}),\ 10.1103/PhysRevB.96.205132}\BibitemShut {NoStop}%
\bibitem [{Note1()}]{Note1}%
  \BibitemOpen
  \bibinfo {note} {In S.I. units, $B$ is the magnetic induction and $B=\mu _0
  H$ with $\mu _0 $ the permeability of the vacuum defines the magnetic field
  intensity $H$, but we adopt the current usage and we call $B$ the magnetic
  field.}\BibitemShut {Stop}%
\bibitem [{Note2()}]{Note2}%
  \BibitemOpen
  \bibinfo {note} {Usually the term effective mass is reserved for the band
  masses, but here we use this term loosely also for second derivatives of the
  dispersion relation in the orbital basis.}\BibitemShut {Stop}%
\bibitem [{\citenamefont {Charlebois}()}]{Charlebois_unpublished}%
  \BibitemOpen
  \bibfield  {author} {\bibinfo {author} {\bibfnamefont {M.}~\bibnamefont
  {Charlebois}},\ }\href@noop {} {\bibinfo  {journal} {Unpublisehd
  calculation}\ }\BibitemShut {NoStop}%
\bibitem [{\citenamefont {Fukuyama}(1969)}]{10.1143/PTP.42.1284}%
  \BibitemOpen
\bibfield  {journal} {  }\bibfield  {author} {\bibinfo {author} {\bibfnamefont
  {H.}~\bibnamefont {Fukuyama}},\ }\href {\doibase 10.1143/PTP.42.1284}
  {\bibfield  {journal} {\bibinfo  {journal} {Progress of Theoretical Physics}\
  }\textbf {\bibinfo {volume} {42}},\ \bibinfo {pages} {1284} (\bibinfo {year}
  {1969})}\BibitemShut {NoStop}%
\bibitem [{\citenamefont {Itoh}(1985)}]{0305-4608-15-8-011}%
  \BibitemOpen
  \bibfield  {author} {\bibinfo {author} {\bibfnamefont {M.}~\bibnamefont
  {Itoh}},\ }\href {http://stacks.iop.org/0305-4608/15/i=8/a=011} {\bibfield
  {journal} {\bibinfo  {journal} {Journal of Physics F: Metal Physics}\
  }\textbf {\bibinfo {volume} {15}},\ \bibinfo {pages} {1715} (\bibinfo {year}
  {1985})}\BibitemShut {NoStop}%
\end{thebibliography}
%

\end{document}